\documentclass{article}

\usepackage{arxiv}

\usepackage[english]{babel}  
\usepackage[utf8x]{inputenc}
\usepackage[T1]{fontenc}

\usepackage{amsmath}
\usepackage{amssymb}
\usepackage{array}
\usepackage{graphicx}
\usepackage[colorinlistoftodos]{todonotes}
\usepackage{algorithm}
\usepackage{algpseudocode}
\usepackage{float}
\usepackage{subfigure}

\usepackage{booktabs}

\usepackage{navigator}

\usepackage{url}
\usepackage{cleveref}

\newcolumntype{C}{>{$}c<{$}} 

\title{Quasi-Monte Carlo methods for calculating derivatives sensitivities on the GPU}

\author{ \anchor{https://orcid.org/0000-0001-6846-6649}{\includegraphics[scale=0.06]{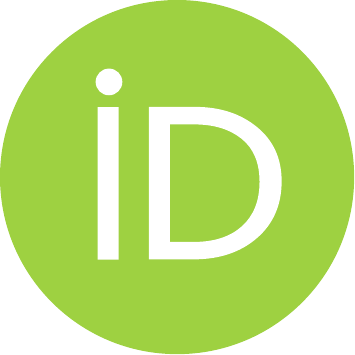}\hspace{1mm}Paul Bilokon} \\
	Department of Computing \\
	Imperial College London \\
	South Kensington Campus \\
	London SW7 2AZ \\
	\texttt{paul.bilokon@imperial.ac.uk} \\
	\And
	\anchor{https://orcid.org/0000-0003-1593-8264}{\includegraphics[scale=0.06]{orcid.pdf}\hspace{1mm}Sergei Kucherenko} \\
	Centre for Process Systems \\
	Imperial College London \\
	South Kensington Campus \\
	London SW7 2AZ \\
	\texttt{s.kucherenko@imperial.ac.uk} \\
	\And
	\anchor{https://orcid.org/0000-0002-9735-3990}{\includegraphics[scale=0.06]{orcid.pdf}\hspace{1mm}Casey Williams} \\
	Department of Computing \\
	Imperial College London \\
	South Kensington Campus \\
	London SW7 2AZ \\
	\texttt{casey.williams19@imperial.ac.uk} \\}

\begin{document}

\maketitle

\begin{abstract}
The calculation of option Greeks is vital for risk management. Traditional pathwise and finite-difference methods work poorly for higher-order Greeks and options with discontinuous payoff functions. The Quasi-Monte Carlo-based conditional pathwise method (QMC-CPW) for options Greeks allows the payoff function of options to be effectively smoothed, allowing for increased efficiency when calculating sensitivities. Also demonstrated in literature is the increased computational speed gained by applying GPUs to highly parallelisable finance problems such as calculating Greeks. We pair QMC-CPW with simulation on the GPU using the CUDA platform. We estimate the delta, vega and gamma Greeks of three exotic options: arithmetic Asian, binary Asian, and lookback. Not only are the benefits of QMC-CPW shown through variance reduction factors of up to $1.0 \times 10^{18}$, but the increased computational speed through usage of the GPU is shown as we achieve speedups over sequential CPU implementations of more than $200$x for our most accurate method.
\end{abstract}

\section{Introduction}

Calculating sensitivities (Greeks) of the value of an option to underlying parameters such as volatility and interest rates, is vital to financial institutions performing risk management and developing hedging strategies. Greeks cannot be observed in the market directly, thus must be calculated from other data.

Traditional finite-difference (FD) methods for calculating Greeks have easy implementations and few restrictions on the form of the payoff function, however they require resimulations which result in estimates with large variances, bias and increased computational effort compared with other methods. 

The pathwise~(PW) method~\cite{glasserman1991gradient} does not require resimulation and provides unbiased estimators; however, it relies on the continuity of the payoff function, therefore it is not applicable to options such as a binary Asian option and cannot be used to calculate most second-order Greeks, where we typically see discontinuity introduced into the function.

The likelihood ratio~(LR) method~\cite{broadie1996estimating} does not require smoothness of the payoff function; although it tends to result in estimates with large variance as it does not use properties of the payoff function.

Introduced by Zhang and Wang~\cite{ZhangConditionalQuasiMonteCarloMethod}, the Quasi-Monte Carlo-based conditional pathwise method~(QMC-CPW) takes a conditional expectation of the payoff function, which results in the discontinuous integrand being smoothed. They also show that the interchange of expectation and differentiation is possible, allowing the estimation of Greeks from the now smooth target function. Through proof of the smoothed payoff being Lipschitz continuous, the PW method is now applicable to provide unbiased estimators. They also show how many options can have infinitely differentiable target functions once the conditional expectation is taken, thus the PW method can be used to calculate second-order Greeks.

GPUs have been discussed extensively in the computational finance literature~\cite{dempster2018hpc}. The highly parallel nature of Monte Carlo simulation for option Greeks lends itself well to the architecture of GPUs and the CUDA architecture. We implement Monte Carlo methods which take advantage of the highly parallel nature of the GPU to gain advantages in speed and efficiency when calculating Greeks.

\subsection{Objectives}

The aim of this work is to apply QMC-CPW to calculate Greeks for options, whilst adapting the implementation to run efficiently on a GPU.

Increased efficiency is not the only aim, but also the broadening of the set of financial products (such as those with discontinuous payoff functions) supported by the algorithm will provide further practical value. As opposed to other solutions developed for the GPU, we will produce unbiased estimates with low variance applicable to options with discontinuous payoff functions and for higher-order Greeks.

\subsection{Challenges}

Adapting algorithms to run on the GPU comes with many restrictions when compared to implementations on the CPU. Memory management and access patterns play a large role in the efficiency and speed when running kernels, so close attention must be paid during implementation to how the on-device memory is used.

CUDA poses further limitations upon the general design of the software such as having separate memory spaces between host and device memory (this has been addressed by unified memory which has been available since toolkit version 6.0). Problems such as these are standard when programming with CUDA and require overhead on the developer's side to ensure code is written in a safe manner.

\subsection{Contributions}

The need for efficient and accurate methods that can be applied to many types of financial options presents an opportunity to use recent methods for Greeks estimation in conjunction with GPUs, and to obtain both an increase in accuracy and speed. Our contributions are as follows:

\begin{enumerate}
    \item Flexible models of products are implemented for the arithmetic Asian, binary Asian and lookback option types. They have a templatised design which allows for minimal reproduction of simulation kernels.
    \item GPU implementation of the Likelihood Ratio method for estimating Greeks, which acts as a baseline to compare variance reduction factors of other methods. All methods estimate the \textit{delta}, \textit{vega}, and \textit{gamma} Greeks.
    \item Implementation of the QMC-CPW method with standard and Quasi-Monte Carlo simulation on both CPU and GPU. CPU implementations are serial and used for comparison of the speedup obtained by using the GPU.
    \item For standard Monte Carlo simulations, antithetic variables are also implemented as a variance reduction technique.
    \item For Quasi-Monte Carlo we perform Brownian bridge construction, which produces Brownian path increments for use in the simulation of the behaviour of the underlying asset, which leads to variance reduction.
    \item We show that the Quasi-Monte Carlo Conditional Pathwise method with Brownian bridge construction (QMC+BB-CPW) is the superior method in terms of accuracy with variance reduction factors of up to $1.0 \times 10^{18}$ and with many in the hundreds of thousands and millions.
    \item We show that using the GPU leads to a massive speedup over the CPU with even the slowest methods being more than $200$x faster.
    \item Finally, it is shown that QMC+BB-CPW implemented on the GPU results in an efficient, accurate, and fast method for calculating first- and higher-order Greeks of options, including those with discontinuous payoff functions. We find Quasi-Monte Carlo takes advantage of the increased smoothness in the integrand following the conditional expectation from CPW, and that the Brownian bridge construction results in further variance reduction.
\end{enumerate}

\section{Preliminaries} \label{sec:Preliminaries}

\subsection{Monte Carlo methods} \label{sec:MonteCarloMethods}

Monte Carlo methods, in the simplest form, rely on repeatedly taking random samples from a set of possible outcomes to determine the fraction of random draws which fall in a given set as an estimate of the set's volume in the probability space~\cite{glasserman2004monte}. As the number of draws increases, the law of large numbers ensures the estimate converges to the true value and information about the magnitude of the error in the estimate can be obtained through the central limit theorem.

Let us use the example from Section~1.1.2 of~\cite{glasserman2004monte}. Suppose we wish to calculate the expected present value of the payoff of a vanilla European call option on a stock under the Black--Scholes model. We can draw samples from the distribution of the terminal stock price \(S(T)\) to calculate the expected value of the payoff~\(E[e^{-rT}(S(T) - K)^+]\). The logarithm of the stock price is normally distributed, so we only need to draw random samples \(Z_i\) from the standard normal distribution to calculate \(S(T)\):

\begin{algorithm}[hbt!]
\caption{Algorithm to estimate expected present value of payoff}\label{alg:EstimateExpectedPayoff}
\begin{algorithmic}[1]
\For{$i \gets 1..n$}
    \State generate $Z_i$
    \State $S_i(T) = S(0)\exp{([r - \frac{1}{2}\sigma^2]T + \sigma \sqrt{T}Z_i)}$
    \State $C_i = e^{-rT}(S_i(T) - K)^+$
\EndFor
\State $\hat{C_n} = (C_1 + \dots + C_n) / n$
\end{algorithmic}
\end{algorithm}

This method can be generalised to calculate payoffs for more exotic path-dependent options and to other problems such as calculating the Greeks of a portfolio of derivatives~\cite{giles2007monte}.

\subsubsection{Pseudorandom number generation} \label{sec:prng}

Randomly sampling from probability distributions using {\emph{pseudorandom number generator~(PRNG)} is at the heart of Monte Carlo. A sequence of random numbers $U_1, U_2, \dots$ should satisfy the following properties: 
\begin{itemize}
    \item \textit{Good randomness properties}
    \begin{enumerate}
        \item Each $U_i$ is uniformly distributed between 0 and 1.
        \item All $U_i$ are mutually independent.
    \end{enumerate}
    The second property is ensured by passing through statistical tests.
    \item \textit{Large period}. The period of a PRNG is the minimum length of the output sequence before any number is repeated. Generators with large periods are key for use in simulation as we wish to draw millions of samples and without a sufficiently large period this would not be possible.
    \item \textit{Speed and efficiency of generation}. As we are generating millions of samples during a single simulation it is necessary for this process to be fast and require little effort computationally.
    \item \textit{Reproducibility}. It is important that using the same seed will result in the same output sequence. This allows us to run simulations multiple times with the same input to verify results.
\end{itemize}

\subsection{Quasi-Monte Carlo} \label{sec:quasimc}
\emph{Quasi-Monte Carlo~(QMC)} uses \emph{low-discrepancy sequences~(LDS)}. which rather than mimic randomness, attempt to generate numbers that are evenly distributed. The advantage of using LDS is the rate at which they converge: while Monte Carlo converges with rate $O(1/\sqrt{n})$, where $n$ is the number of paths, QMC has asymptotic convergence rate close to $O(1/n)$. However, at the practically achievable $n$ QMC may have a dependence on the dimension of the problem. With many financial problems having high dimension due to large numbers of risk factors, time steps per path, and the number of paths simulated, it is not guaranteed that QMC has greater performance over Monte Carlo. This has been addressed through a number of techniques such as \emph{variance reduction}~\cite{allen2011variance, WANGvariancereduction}, and the concept of \emph{effective dimension}~\cite{caflisch_1998, WANGeffectivedim, bianchetti2015pricing} explains the success of QMC even for problems of high dimension.

Consider the problem of numerical integration over the unit hypercube $[0, 1)^d$. We want to calculate
\begin{equation} \label{eqn:UnitHypercubeIntegral}
    \mathbb{E}[f(U_1, \dots, U_d)] = \int_{[0,1)^d} f(x) dx,
\end{equation}
This integral is approximated by
\begin{equation} \label{eqn:ApproxUnitHypercubeIntegral}
    \int_{[0,1)^d} f(x) dx \approx \frac{1}{n} \sum_{i = 1}^n{f(x_i)}.
\end{equation}

To calculate this value using Monte Carlo, we construct a vector $(U_1, \dots, U_d)$ from i.i.d. sequence of points uniformly distributed on the unit hypercube. In the case of QMC we generate vectors of $d$-dimensional LDS points. The most commonly used LDS in finance are Sobol' sequences~\cite{sobol1967distribution}.

\subsubsection{Van der Corput sequences} \label{sec:VanDerCorputSequences}

Sobol' sequences are based on permutations of the Van der Corput sequences in base 2. This sequence is a specific class of LDS in one dimension and is the core of many multidimensional constructions.

Every positive integer $k$ has what is known as its base-$b$ representation such that
\begin{equation} \label{eqn:BaseBKRepresentation}
    k = \sum_{j=0}^{\infty}{a_j(k)b^j},
\end{equation}
where $b \ge 2$ and finitely many of the coefficients $a_j(k)$ are not equal to zero and in $\{0,1\dots,b-1\}$. The \emph{radical inverse function} $\psi_b$ is a mapping of each $k$ to $[0,1)$ and is given as
\begin{equation} \label{eqn:RadicalInverseFunction}
    \psi_b(k) = \sum_{j=0}^{\infty}{\frac{a_j(k)}{b^{j+1}}}.
\end{equation}

\subsubsection{Sobol' sequences} \label{sec:SobolSequences}

The Sobol' points start from the Van der Corput sequence in base $2$, and the coordinates of a $d$-dimensional sequence come from permutations of sections of the Van der Corput sequence. These permutations result from the product of binary expansions of consecutive integers with a set of generator matrices, one for each dimension. A generator matrix $\boldsymbol{G}$ has columns of binary expansions of a set \emph{direction numbers} $g_1,\dots,g_r$ with elements equal to $0$ or $1$. The value $r$ represents the number of terms in the binary expansion of $k$ and can be arbitrarily large. Let $(a_0(k),\dots,a_{r-1}(k))^\top$ represent the vector of coefficients of the binary representation of $k$ such that
\begin{equation} \label{eqn:GeneratorMatrixExpression}
    \begin{pmatrix}
    y_1(k) \\
    y_2(k) \\
    \vdots \\
    y_r(k)
    \end{pmatrix}
    = \boldsymbol{G} 
    \begin{pmatrix}
    a_0(k) \\
    a_1(k) \\
    \vdots \\
    a_{r-1}(k)
    \end{pmatrix}
    \mod{2},
\end{equation}
and $y_1(k),\dots,y_r(k)$ are the coefficients of the binary expansion of the $k$th point in the sequence. This gives the $k$th point as:
\begin{equation*}
    x_k = \frac{y_1(k)}{2} + \frac{y_2(k)}{4} + \dots + \frac{y_r(k)}{2^r}.
\end{equation*}

The generator matrix~$\boldsymbol{G}$ is upper triangular and the special case where it is the identity matrix results in the Van der Corput sequence in base $2$. We can perform~(\ref{eqn:GeneratorMatrixExpression}) in a computer implementation through a bitwise XOR operation, giving us the computer representation of $x_k$ as
\begin{equation*}
    a_0(k)g_1 \oplus a_1(k)g_2 \oplus \dots \oplus a_{r-1}(k)g_r,
\end{equation*}
where $\oplus$ is the bitwise XOR operator.

The core of the Sobol' method are the generator matrices~$\boldsymbol{G}$ and their direction numbers~$g_j$. As previously mentioned, we require $d$ sets of direction numbers to produce a $d$-dimensional sequence. The method begins by selecting a \emph{primitive polynomial} over binary arithmetic. The polynomial
\begin{equation} \label{eqn:SobolPrimitivePolynomial}
    x^q + c_1x^{q-1} + \dots + c_{q-1}x + 1,
\end{equation}
has coefficients $c_i$ in $\{0,1\}$ and satisfies two properties~\cite{glasserman2004monte}:

\begin{itemize}
    \item it cannot be factored;
    \item the smallest power $p$ for which the polynomial divides $x^p + 1$ is $p = 2^q - 1$.
\end{itemize}

The primitive polynomial in~(\ref{eqn:SobolPrimitivePolynomial}) defines a recurrence relation
\begin{equation} \label{eqn:SobolRecurrenceRelation}
    m_j = 2c_1m_{j-1} \oplus 2^2c_2m_{j-2} \oplus \dots \oplus 2^{q-1}c_{q-1}m_{j-q+1} \oplus 2^qm_{j-q} \oplus m_{j-q},
\end{equation}
where the $m_j$ are integers. We define the directions numbers as
\begin{equation*}
    g_j = \frac{m_j}{2^j}.
\end{equation*}

Of course, to fully define the direction numbers we need initial values for $m_1,\dots,m_q$. It is enough to set each initialising $m_j$ to be an odd integer less than $2^j$, which ensures that all following $m_j$ as defined by~(\ref{eqn:SobolRecurrenceRelation}) also share this property. From this, each $g_j$ will be strictly between $0$ and $1$.

So, to construct a sequence we take the primitive polynomial and use the recurrence relation~(\ref{eqn:SobolRecurrenceRelation}) with some initial $m_j$. We then calculate the corresponding direction numbers~$g_j$ by dividing by $2^j$ (or performing a binary shift of the binary point $j$ places to the left). Then with these direction numbers we construct the generator matrix $\boldsymbol{G}$. With this generator matrix we take a vector $\boldsymbol{a}(k)$ of binary coefficients of $k$ and perform the operation in~(\ref{eqn:GeneratorMatrixExpression}) to give us the coefficients of a binary fraction, from which we obtain $x_k$.

There has been much research on choosing initial direction numbers, and also more efficient construction implementation (e.g. ~\cite{sobol2011construction}), which we will not go into further detail about.

\subsubsection{Scrambled Sobol'} \label{sec:ScrambledSobol}

As we are choosing points deterministically we are unable to measure error through a confidence interval. Randomised QMC points allow us to calculate this error. One method for producing randomised QMC points is known as \emph{scrambling}. Introduced by Owen and further developed in~\cite{OwenScrambling}, scrambling is a technique that permutes each digit of a $b$-ary expansion, where the permutation applied to the $j$th digit is dependent on the preceding $j-1$ digits. Scrambling can be described by taking each coordinate, partitioning the unit interval into $b$ subintervals of length $1/b$, and then randomly permuting those subintervals. Then, further partition each subinterval into $b$ subintervals of length $1/b^2$ and permute those, and so on. At the $j$th step, we have $b^{j-1}$ partitions, each of which consist of $b$ intervals, and each is permuted randomly and independently.

\subsection{Graphics Processing Units and CUDA}

The \emph{Graphics Processing Unit (GPU)}~\cite{dempster2018hpc} has seen widespread adoption in computational finance due to its highly parallel architecture designed for increased computational throughput. When NVIDIA released CUDA~\cite{cudazone} in 2007 it enabled more ``general-purpose'' usage of the previously graphics-focused applications of GPUs.

\subsubsection{CUDA architecture} \label{sec:cudaarch}

The CUDA architecture allows each and every arithmetic logic unit~(ALU) on the chip to be marshaled by a program~\cite{sanders2010cuda}. It is implemented by organising the GPU into a collection of \emph{streaming multiprocessors}, which operate following the \emph{Single-Instruction-Multiple-Thread~(SIMT)} paradigm. Because of the intended usage for general-purpose computation, CUDA allows for arbitrary read and write access to memory and the software-managed cache known as \emph{shared memory}.

From a software perspective, the CUDA architecture allows for \emph{kernels} to be ran in parallel across a \emph{grid}. This grid is composed of multiple \emph{blocks}, each of which contains a collection of threads which all run the program defined by some launched kernel. Both blocks and grids can have up to three dimensions each, and CUDA provides useful syntax for indexing into them. In hardware, the threads inside of a block are grouped into sets of 32 threads known as a \emph{warp}, where all threads inside the same warp execute the same instruction.

Each thread has its own local memory and registers, and threads in the same block have access to the on-chip shared memory of that block. This is often how threads within a block communicate with each other while maintaining high performance. 

NVIDIA have developed a toolkit for CUDA~\cite{cudatoolkitdocumentation} which contains the compiler, highly parallel implementations of mathematical libraries (such as cuBLAS, cuRAND and cuFFT), and a host of other useful tools like a debugger and memory checker.

\subsubsection{Practical implementation considerations}

There are many considerations one must take into account when implementing algorithms on a GPU. Most notably, the limited size of on-chip caches in comparison to the relatively large size of global memory. For financial problems with high dimensions (such as Monte Carlo simulations of many paths or many assets) shared memory will quickly become a limiting factor to the speed of an implementation. This is because reading from global memory is roughly $100$x slower than loading directly from shared memory. This limitation has been addressed in the literature and a few common design patterns have arisen such as pre-computation of values shared between threads, merging of kernels to avoid redundant data transfers, and using coalesced reads and writes. See~\cite{dixon2012monte} for an example of how problem reformation can lead to large speed-ups and see~\cite{BrodtkorbGPU} for further discussion of GPU programming strategies.
\section{Background} \label{sec:Background}

\subsection{Calculating Greeks}

Calculating price sensitivities (Greeks) is vital for risk management and hedging. The calculation of Greeks requires significant computational effort when compared to that of determining derivative prices, thus efficient implementation of algorithms for obtaining sensitivities is key for financial institutions.

\subsubsection{Finite-difference method}

The simplest method for obtaining sensitivities is based on the \emph{finite-difference} approach. Within the Monte Carlo framework this involves running multiple simulations of a pricing routine over a range of values of input parameters. For example, determining the delta of a call option would involve running simulations for different values of the underlying price and observing the changes in the resulting option price. To obtain the derivative of an options price with respect to input parameter $\theta$ we would estimate 
\begin{equation*}
    \frac{\partial V(t, \theta)}{\partial \theta} \approx \frac{V(t, \theta + h) - V(t, \theta)}{h},
\end{equation*}
where $V(t, \theta)$ is the value of the payoff of the option at time $t$ and some small $h \in \mathbb{R}^+$ known as the ``bump size''.

The finite-difference method is intuitive and easy to implement; however, it requires significantly higher computation time as the number of input parameters grows and suffers from poor bias and variance properties.

\subsubsection{Pathwise method} \label{sec:PathwiseMethod}

An alternative to finite-difference is the \emph{pathwise method}. Developed by Glasserman~\cite{glasserman1991gradient} and explained further by Broadie and Glasserman~\cite{broadie1996estimating}, the pathwise method has two main benefits: increased computational speed and unbiased estimates. To explain the pathwise method, let us consider the calculation of the delta of a vanilla European call option on a stock that satisfies the Black--Scholes SDE. Let $Y$ denote the present value of the payoff
\begin{equation*}
    Y = e^{-rT}[S(T) - K]^+.
\end{equation*}
Applying the chain rule we obtain
\begin{equation} \label{eqn:ChainRuleOptionDelta}
    \frac{\partial Y}{\partial S(0)} = \frac{\partial Y}{\partial S(T)} \frac{\partial S(T)}{\partial S(0)}.
\end{equation}

Observe that
\begin{equation*}
    S(T) = S(0) \exp\left(\left[r - \frac{1}{2}\sigma^2\right] T + \sigma W(T)\right)
\end{equation*}
is linear in $S(0)$ and so $\partial S(T)/ \partial S(0) = S(T)/S(0)$. We have $\partial Y/ \partial S(T) = e^{-rT}\textbf{1}\{S(T) > K\}$, combining the two gives us the pathwise estimator for the delta
\begin{equation} \label{eqn:PathwiseDeltaEstimate}
    \frac{\partial Y}{\partial S(0)} = e^{-rT} \frac{S(T)}{S(0)} \textbf{1}\{S(T) > K\}.
\end{equation}

We can obtain other first-order and higher-order derivatives through similar means. It can be seen that~(\ref{eqn:PathwiseDeltaEstimate}) is easily evaluated and has been shown to be an unbiased estimator~\cite{broadie1996estimating}. The method can also be applied to path-dependent options and provides a lot of practical value for options with no closed-form solution (such as Asian options). Further, as many of the factors used in calculating an option's price are present in the pathwise estimators, little effort is required to add them to an existing pricing implementation.

To provide context of how pathwise is used within Monte Carlo, let us consider calculating the delta of a derivative security with multiple underlying assets and payoff function $f$. We model the evolution of a stock price where $W$ is a $d$-dimensional Brownian motion, and we are approximating the price using an Euler scheme with timestep $h = T/N$. We can write the Euler approximation at time $nh$ as follows:
\begin{equation} \label{eqn:EulerApproximation}
    \hat{S}(n + 1) = \hat{S}(n) + a(\hat{S}(n))h + b(\hat{S}(n)) Z(n+1) \sqrt{h}, \quad \hat{S}(0) = S(0), 
\end{equation}
with $a(\cdot) \in \mathbb{R}^m$, $b(\cdot) \in \mathbb{R}^{m \times d}$ and $Z(1), Z(2), \dots$ are $d$-dimensional standard normal random vectors. (\ref{eqn:EulerApproximation})~then takes the form
\begin{equation} \label{eqn:EulerMatrixForm}
    \hat{S}(n + 1) = F_n(\hat{S}(n)),
\end{equation}
with $F_n$ a matrix transformation $\mathbb{R}^m \to \mathbb{R}^m$. Then we can perform similar operations as in~(\ref{eqn:ChainRuleOptionDelta}), we obtain the pathwise estimate of the delta
\begin{equation} \label{eqn:MatrixPathwiseDeltaEstimate}
    \sum_{i=1}^m \frac{\partial f(\hat{S}(N)}{\partial \hat{S}_i(N)} \Delta_{ij}(N)
\end{equation}
with
\begin{equation*}
    \Delta_{ij}(n) = \frac{\partial \hat{S}_i(n)}{\partial \hat{S}_j(0)}, \quad i,j = 1, \dots, m.
\end{equation*}

This can be written as a matrix recursion
\begin{equation} \label{eqn:DeltaMatrixRecursion}
    \Delta (n + 1) = G(n) \Delta (n), \quad \Delta (0) = I,
\end{equation}
where $G(n)$ represents the derivative of the transformation $F_n$ and $\Delta (n)$ is the $m \times m$ matrix with entries $\Delta_{ij}(n)$.

There are some limitations to the pathwise method, namely the payoff function must be Lipschitz continuous but there exist other methods to overcome this problem such as smoothing the payoff function, using the Likelihood Ratio Method (LRM)~\ref{sec:LikelihoodRatioMethod} or an alternative form of Monte Carlo simulation such as ``Vibrato'' Monte Carlo~\cite{giles2009vibrato}.

\subsubsection{Likelihood ratio method} \label{sec:LikelihoodRatioMethod}

Rather than view the final state of a stock price as a random variable, we can look from the perspective of a probability distribution~\cite{broadie1996estimating}. For an option with payoff function $Y = f(S(T))$ and underlying satisfying the Black--Scholes SDE, such that the payoff is expressed as a function of a random vector $X = (X_1,\dots,X_d)$, its value can be written as
\begin{equation} \label{eqn:OptionValueProbabilityForm}
    V = \mathbb{E}[f(Y)] = \int f(x) g_\theta (x) dx,
\end{equation}
where $g_\theta$ is probability density function of $X$. Supposing that the interchange of order between integration and differentiation holds, we can take the derivative of~(\ref{eqn:OptionValueProbabilityForm}) with respect to an input parameter~$\theta$ to obtain the likelihood ratio estimator
\begin{equation} \label{eqn:LRUnbiasedEstimate}
    \frac{\partial V}{\partial \theta} = \int f(x) \frac{g^\prime_\theta(x)}{g_\theta(x)} g_\theta(x)dx = E\left[f(X) \frac{g^\prime_\theta(X)}{g_\theta(X)} \right].
\end{equation}

As probability densities are generally continuous, we can apply the \emph{likelihood ration method~(LRM)} to calculate Greeks for derivatives with discontinuous payoff functions and, as with the pathwise method, it works well for path-dependent options.

A weakness of LRM lies in its $O(h^{-1})$ estimator variance where $h$ is the timestep for the path discretisation in simulation.

\subsection{Monte Carlo methods}

Monte Carlo simulation is an essential tool in computational finance for calculating prices of derivatives and their sensitivities to input parameters, commonly known as the ``Greeks''. The application of Monte Carlo simulation to pricing derivatives was first developed by Boyle in 1977~\cite{BOYLE1977323} and has shown to be an efficient method for high-dimensional problems. The ease of implementation and intuitiveness behind Monte Carlo have continued to make it a key approach for many problems in computational finance~\cite{glasserman2004monte}.

Following Boyle's seminal paper, application of Monte Carlo methods to many problems in finance and the acceleration of implementations became a focus in literature. For a review of early Monte Carlo methods and their use for calculating derivatives prices see~\cite{boyle1997monte}.

Broadie and Glasserman~\cite{broadie1996estimating} develop two techniques which allow for increased computational speed over the traditional finite-difference method when calculating derivative sensitivities through Monte Carlo simulation. The basics of these two methods are detailed in Sections~\ref{sec:PathwiseMethod} and~\ref{sec:LikelihoodRatioMethod}. These ``direct methods'' not only speed up simulation but provide \emph{unbiased estimators} for sensitivities, unlike finite-difference, and work for path-dependant options.

The issue of discontinuous payoff functions has been discussed extensively in literature and still continues to be a popular topic. Giles presents the ``Vibrato'' Monte Carlo method~\cite{giles2009vibrato} which combines the adjoint pathwise approach for the stochastic path evolution, with LRM for evaluation of the payoff function. He shows that, when the payoff function is discontinuous, the resulting estimator has variance $O(h^{-1/2})$, where $h$ is the timestep for the path discretisation, and $O(1)$ when the payoff is continuous. The numerical results presented show its superior efficiency when compared to standard LRM.

\subsubsection{GPU implementations}

There are several properties of Monte Carlo which make it attractive for an implementation with high parallelism, thus in recent years much work has been done on using GPUs to accelerate these simulations.

Dixon \textit{et al.}~\cite{dixon2012monte} show that Monte Carlo is well suited to implementation on a high performance GPU and discuss methods for accelerating Value-at-Risk estimation through several key implementation techniques. More recently, the techniques discussed in Section~\ref{sec:PathwiseMethod} paired with Algorithmic Adjoint Differentiation~(AAD) have also seen implementation on the GPU~\cite{savickas2014super} and have shown speed-ups of over 10 times when compared to traditional finite difference methods on GPUs, and more than 70 times when compared to multi-core CPU implementations.

\subsection{Variance reduction techniques}

Boyle \textit{et al.}~\cite{boyle1997monte} discuss variance reduction techniques and show that their application reduces the error in estimates, thus increasing the efficiency of Monte Carlo simulation. In its simplest form, the argument for variance reduction techniques is to increase \emph{efficiency}. If we have two (unbiased) Monte Carlo estimates for parameter $\theta$, denoted by $\{ \hat{\theta}_i^{(1)}, i = 1, 2, \dots \}$ and $\{ \hat{\theta}_i^{(2)}, i = 1, 2, \dots \}$, with $b^{(j)}, j = 1, 2$, the computational work required to generate one replication of $\hat{\theta}^{(j)}$, then we would choose estimator $1$ over $2$ if
\begin{equation} \label{eqn:EstimatorEfficiencyComparison}
    \sigma_1^2 b_1 < \sigma_2^2 b_2,
\end{equation}
where $\sigma_j^2$ is the variance of the estimator $\hat{\theta}^{(j)}$. We can take the product of variance and computational work to be a measure of the efficiency, thus use~(\ref{eqn:EstimatorEfficiencyComparison}) as a way to compare multiple Monte Carlo estimators. We briefly detail some of the common techniques to reduce variance in the following sections. For further explanation, the reader is referred to~\cite{glasserman2004monte, boyle1997monte}.

\subsection{Quasi-Monte Carlo-based conditional pathwise method} \label{sec:qmc-cpwmethod}

As an extension to the pathwise method described in Section~\ref{sec:PathwiseMethod}, Zhang and Wang~\cite{ZhangConditionalQuasiMonteCarloMethod} introduce the Quasi-Monte Carlo-based conditional pathwise method.

Let us denote the discounted payoff of an option $g(\theta,\boldsymbol{x})$ as
\begin{equation} \label{eqn:QmcCpwDiscountedPayoff}
    g(\theta,\boldsymbol{x}) = h(\theta,\boldsymbol{x})\boldsymbol{1}\{p(\theta,\boldsymbol{x}) > 0\},
\end{equation}
where $h(\theta,\boldsymbol{x})$ and $p(\theta,\boldsymbol{x})$ are continuous functions of $\theta$ and $\boldsymbol{x}$. The function $p(\theta,\boldsymbol{x})$ is said to satisfy the \emph{variable separation condition} if
\begin{equation} \label{eqn:QmcCpwVariableSeparation}
    \boldsymbol{1}\{p(\theta,\boldsymbol{x}) > 0\} = \boldsymbol{1}\{\psi_d(\theta,\boldsymbol{z}) < x_j < \psi_u(\theta,\boldsymbol{z})\},
\end{equation}
for some variable $x_j$, where $\psi_d(\theta,\boldsymbol{z})$ and $\psi_u(\theta,\boldsymbol{z})$ are functions of $\theta$ and $\boldsymbol{z}$ where
\begin{equation*}
    \boldsymbol{z} = (x_1,\dots,x_{j-1},x_{j+1},\dots,x_d)^\top.
\end{equation*}

Then, if $p(\theta,\boldsymbol{z})$ satisfies (\ref{eqn:QmcCpwVariableSeparation}), the discounted payoff in (\ref{eqn:QmcCpwDiscountedPayoff}) can be written as
\begin{equation} \label{eqn:QmcCpwPayoffSeparated}
    g(\theta,\boldsymbol{x}) = h(\theta,\boldsymbol{x})\boldsymbol{1}\{\psi_d(\theta,\boldsymbol{z}) < x_j < \psi_u(\theta,\boldsymbol{z})\}.
\end{equation}

Using Fubini's theorem, the discounted payoff $g(\theta,\boldsymbol{x})$ is first integrated with respect to $x_j$, such that we can write the price of the option as $\mathbb{E}[G(\theta,\boldsymbol{z})]$ where 
\begin{equation} \label{eqn:QmcCpwNewTargetDefinition}
    \mathbb{E}[g(\theta,\boldsymbol{x})|\boldsymbol{z}] = \int_{\psi_d}^{\psi_u}{h(\theta,\boldsymbol{x})\rho_j(x_j)dx_j\boldsymbol{1}\{\psi_d(\theta,\boldsymbol{z}) < \psi_u(\theta,\boldsymbol{z})\}} = G(\theta, \boldsymbol{z}),
\end{equation}

and we assume can be found analytically. We can then interchange expectation and differentiation (as with the pathwise method) to obtain estimates of Greeks.

Zhang and Wang show that the discounted payoffs of many options under the Black-Scholes model satisfy the variable separation condition. Following proof that the interchange of expectation and differentiation is valid, and defining the new target function $G(\theta,z)$ as the expectation of the discounted payoff~(\ref{eqn:QmcCpwPayoffSeparated}) conditioned on $z$, it is shown that the new estimate for the sensitivity of the payoff to parameter $\theta$ is unbiased even when the original payoff~(\ref{eqn:QmcCpwDiscountedPayoff}) is not continuous.

It can be easily shown that $G(\theta,z)$ is a continuous function of $z$ (demonstrated by Theorem~A.1 in Appendix~1 of~\cite{ZhangConditionalQuasiMonteCarloMethod}). Using the idea of variable separation and taking the conditional expectation, the new target function is smoother than the original payoff function, therefore benefits from QMC in practice.

\subsubsection{Simulating stock price for variable separation} \label{sec:VariableSeparationPathSimulation}

In order to understand the example in Section~\ref{sec:BaDeltaExample} we must first understand how to simulate the underlying asset's price movement such that variable separation is possible. Here we give a brief overview of the method described in~\cite{ZhangConditionalQuasiMonteCarloMethod}. Following on from the Black--Scholes model, let
\begin{equation} \label{eqn:Stilde1}
\begin{aligned}
    \widetilde{S}(t_j) &= S(0)\exp{(\omega(t_j - t_1) + \sigma(W(t_j) - W(t_1)))} \\
    &= S(0)\exp{(\omega(t_j - t_1) + \sigma\widetilde{W}(t_j - t_1))},
\end{aligned}
\end{equation}

where $\widetilde{W}(t) = W(t + t_1) - W(t_1)$. It is easy to see that $\widetilde{W}(t)$ is also a standard Brownian motion. From~(\ref{eqn:Stilde1}) we have
\begin{equation}
    S(t_j) = \widetilde{S}(t_j)\exp{(\omega t_1 + \sigma W(t_1))}.
\end{equation}

Let $\widetilde{\boldsymbol{W}} = (\widetilde{W}(t_2 - t_1),\dots,\widetilde{W}(t_d - t_1))^\top$ and note that $W(t_1)$ and $\widetilde{\boldsymbol{W}}$ are independent and normally distributed so we are able to generate them as follows
\begin{equation}
    W(t_1) = \sqrt{t_1}x_1, \quad x_1 \sim N(0, 1),
\end{equation}
\begin{equation} \label{eqn:WboldTilde}
    \widetilde{\boldsymbol{W}} = \boldsymbol{Az}, \quad \boldsymbol{z} \sim N(\boldsymbol{0}_{d-1}, \boldsymbol{I}_{d-1}),
\end{equation}
where $\boldsymbol{z} = (x_2,\dots,x_d)^\top$. $\boldsymbol{0}_{d-1}$ is a $d-1$ dimensional zero column vector and $\boldsymbol{I}_{d-1}$ is $d-1$ dimensional identity matrix. The $(d-1) \times (d-1)$ matrix $\boldsymbol{A}$ satisfies $\boldsymbol{AA}^\top = \boldsymbol{\Sigma}$ where
\begin{equation*}
    \boldsymbol{\Sigma} = 
    \begin{pmatrix}
    t_2 - t_1 & t_2 - t_1 & \dots & t_2 - t_1 \\
    t_2 - t_1 & t_3 - t_1 & \dots & t_3 - t_1 \\
    \vdots & \vdots & \ddots & \vdots \\
    t_2 - t_1 & t_3 - t_1 & \dots & t_d - t_1 \\
    \end{pmatrix}
\end{equation*}

There exists much literature on the choice of the matrix $\boldsymbol{A}$, and a good path generation method can reduce the error of the estimates produced.

From (\ref{eqn:Stilde1})-(\ref{eqn:WboldTilde}) we obtain
\begin{equation} \label{eqn:StjFinal}
    S(t_j) = \widetilde{S}(t_j)\exp{(\omega t_1 + \sigma \sqrt{t_1}x_1)}.
\end{equation}

It is clear to see that the stock price $S(t_j)$ at time $t_j$ is a product of the exponential term, and $\widetilde{S}(t_j)$, which are functions of $x_1$ and $\boldsymbol{z}$ respectively. This fact allows many options to satisfy the variable separation conditions, thus we are able to take the conditional expectation to find $G(\theta,\boldsymbol{z})$ and differentiate with respect to the parameter of interest.

\subsubsection{Example: Binary Asian delta by QMC-CPW} \label{sec:BaDeltaExample}

As an example, let us consider the calculation of the delta of a binary Asian option with discounted payoff
\begin{equation}
    g(\theta,\boldsymbol{x}) = e^{-rT}\boldsymbol{1}\{S_A > K\}
\end{equation}
where $S_A$ is the arithmetic average of the stock price $S(t_j)$ and $K$ is the strike. Then from the definition of $S(t_j)$ we obtain
\begin{equation} \label{eqn:SADefinition}
    S_A = \exp{(\omega t_1 + \sigma \sqrt{t_1}x_1)} \frac{1}{d} \sum_{j=1}^{d}{\widetilde{S}(t_j)} = \widetilde{S}_A \exp{(\omega t_1 + \sigma \sqrt{t_1}x_1)},
\end{equation}
with $\widetilde{S}_A$ as the arithmetic average of $\widetilde{S}(t_j)$ for $j = 1,\dots,d$. From (\ref{eqn:SADefinition}) we can see that
\begin{equation*}
    \{S_A > K\} = \{x_1 > \psi_d\},
\end{equation*}
where
\begin{equation*}
    \psi_d = \frac{\ln{K} - \ln{\widetilde{S}_A} - \omega t_1}{\sigma \sqrt{t_1}}
\end{equation*}
and is a function of $\boldsymbol{z}$ only. From this we have achieved the variable separation form listed in~(\ref{eqn:QmcCpwVariableSeparation}). We are now able to calculate the analytical solution of $G(\theta, \boldsymbol{z})$:
\begin{equation} \label{eqn:GAnalyticalSolution}
    \begin{aligned}
    E[g(\theta,\boldsymbol{x})|\boldsymbol{z}] &= \int_{-\infty}^{+\infty}{e^{-rT}\boldsymbol{1}\{S_A > K\}\phi(x_1)dx_1} \\
    &= \int_{-\infty}^{+\infty}{e^{-rT}\boldsymbol{1}\{x_1 > \psi_d\}\phi(x_1)dx_1} \\
    &= \int_{\psi_d}^{+\infty}{e^{-rT}\phi(x_1)dx_1} \\
    &= e^{-rT}[1 - \Phi(\psi_d)] = G(\theta,\boldsymbol{z}).
    \end{aligned}
\end{equation}

Here $\phi(x)$ and $\Phi(x)$ denote the normal density function and the normal cumulative distribution function respectively. The proof of validity of interchange of expectation and differentiation will not be shown here and the reader is referred to~\cite{ZhangConditionalQuasiMonteCarloMethod} for further details.

By differentiating~(\ref{eqn:GAnalyticalSolution}) with respect to the initial stock price $S(0)$ we obtain the conditional pathwise estimate for the delta:
\begin{equation*}
    \begin{aligned}
    \frac{ \partial G }{ \partial S(0)} &= -e^{-rT} \phi(\psi_d) \frac{\partial \psi_d}{ \partial S(0)} \\
    &= e^{-rT} \phi(\psi_d) \frac{1}{\sigma \sqrt{t_1}} \frac{1}{\widetilde{S}_A} \frac{\widetilde{S}_A}{S(0)} \\
    &= \frac{e^{-rT}}{S(0)\sigma \sqrt{t_1}}\phi(\psi_d).
    \end{aligned}
\end{equation*}

\subsection{Related work}

As previously mentioned, the QMC-CPW method~\cite{ZhangConditionalQuasiMonteCarloMethod} can be viewed as an extension to the PW method developed by Glasserman~\cite{glasserman1991gradient}. In their paper, Zhang and Wang consider the relationship of QMC-CPW with current methods other than traditional PW. They show the similarity in the estimates produced by Lyuu and Teng in their LT method~\cite{LyuuYuh-Dauh2010UaeG} despite approaching the problem from different perspectives.

The idea of conditional Monte Carlo is not new, however, and has been covered widely. Boyle and Glasserman~\cite{boyle1997monte} discuss how the technique exploits the variance reducing property of conditional expectation such that for two random variables $X$ and $Y$, $\text{Var}[\mathbb{E}[X|Y]] \le \text{Var}[X]$, typically with a strict inequality except in a few trivial cases. The variance reduction is effectively achieved because we are doing part of the integration analytically by conditioning, leaving a simpler task for Monte Carlo simulation. Glasserman also discusses taking conditional expectation in order to smooth the discounted payoff. In Section~7.2 of~\cite{glasserman2004monte} we see the idea of conditional expectation applied to a digital payoff such that the traditional PW method can be used to obtain and unbiased estimate for the delta (which is not possible with PW alone).
\section{Implementation} \label{sec:Implementation}

The design considerations and their reflecting implementations are detailed in this section.

\subsection{Path simulation}

To simulate a path following a Brownian motion, as in~(\ref{eqn:Stilde1})--(\ref{eqn:StjFinal}), we must generate and consume normal random variables~$Z_i$. Our goal is to improve efficiency and speed when calculating Greeks and so we are not concerned with the performance when generating random variables. The basics of random number generation are discussed in sections~\ref{sec:prng} through~\ref{sec:ScrambledSobol}. There exist many libraries for random number generation and we choose to use cuRAND~\cite{curand} due to it being part of the CUDA toolkit.

To utilise the highly parallel nature of the GPU, each thread will be responsible for the simulation of one path. This requires each thread to have access to it's own distinct set of random variables and a place to store the results from path simulation. The loading and storing of these values is of key importance during the simulation. Due to the number of random variables required we store the arrays in \emph{global memory} which is a slower, but larger, type of memory available in the CUDA architecture. The access pattern to global memory can have a huge impact on the performance of a kernel. Here, we detail the concept of \emph{coalesced memory accesses}. As discussed in~\ref{sec:cudaarch} threads are arranged into groups of 32 known as a \emph{warp}. Accesses to global memory in CUDA are coalesced such that 32-, 64- and 128-byte accesses are loaded in a single transaction, shown in Figure~\ref{fig:coalescedaccess}. In our implementation, each block contains $64$ threads, so at each timestep two warps will load their random variables in just two memory transactions.

\begin{figure}[h]
    \centering
    \includegraphics[width=0.8\textwidth]{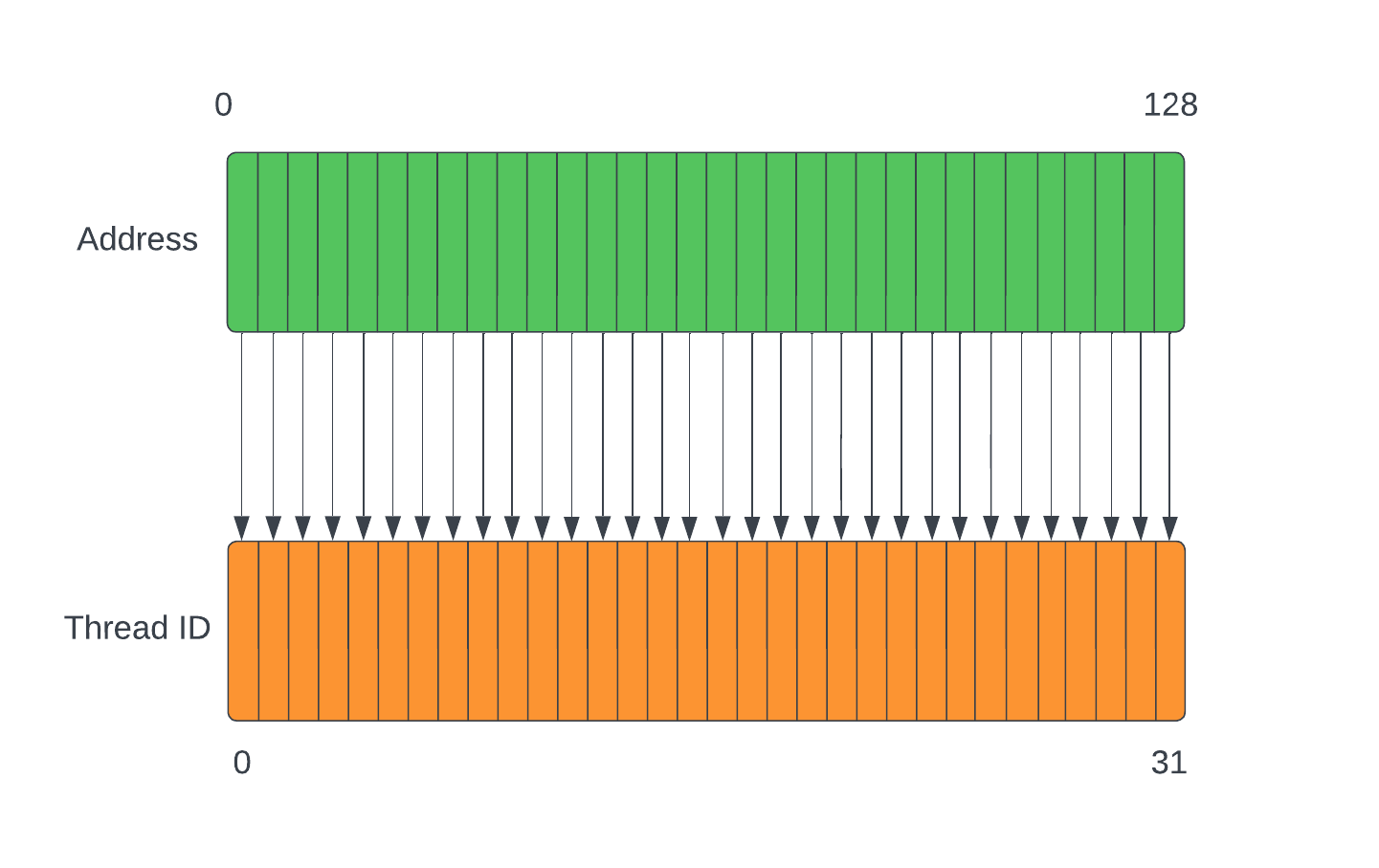}
    \caption{Coalesced memory access where a warp of 32 threads loads 128-bytes in a single transaction. Inspired by: \protect\url{https://cvw.cac.cornell.edu/gpu/coalesced}}
    \label{fig:coalescedaccess}
\end{figure}

Therefore it is extremely important that we load random variables in a way that minimises the number of transactions (due to the much slower global memory). If we were to arrange the accesses such that each thread were to load $N$ contiguous random variables from memory during path simulation, at each step we would have to sequentially perform a separate memory transaction for each thread. This can incur costs of a lot more than 10x when compared to coalesced accesses. As such, we access random variables such that a single transaction satisfies a whole warp.

To perform path generation, we use two types of random number generator from cuRAND: \texttt{CURAND\_RNG\_PSEUDO\_DEFAULT} and \texttt{CURAND\_RNG\_QUASI\_SCRAMBLED\_SOBOL32}. Due to the nature of low-discrepancy sequences we must specify a dimension for the Sobol' generator, we use the number of timesteps in a simulation. We have to pay close attention to the dimensions when using the random variables from the quasi generator as the simulation of each timestep must be independent from each other, thus we must use a random variable from a different dimension. By default, the cuRAND Sobol' generator will output $N/d$ numbers from dimension $1$, followed by $N/d$ from dimension $2$ when generating $N$ variables in $d$ dimensions. The ordering of dimensions is not well spatially-located so we choose to transform the ordering so that coalesced memory access with a smaller stride are possible. Algorithm~\ref{alg:QuasiRandomNumbersTransformation} demonstrates this transformation.

\begin{algorithm}[hbt!]
\caption{Transformation of quasi-random variables from $N*\text{PATHS}/d$ of each dimension to BLOCK\_SIZE of each dimension repeated, where $N$ is the number of timesteps.}\label{alg:QuasiRandomNumbersTransformation}
\begin{algorithmic}[1]
\State $\text{d\_z[PATHS*N]}$ \Comment{Output array} 
\State $\text{temp\_z[PATHS*N]}$ \Comment{Input array of random numbers}
\State $\text{desired\_idx} \gets \text{threadIdx.x} + \text{N} * \text{blockIdx.x} * \text{blockDim.x}$
\State $\text{temp\_idx} \gets \text{threadIdx.x} + \text{blockIdx.x} * \text{blockDim.x}$
\For{$i \gets 0..N-1$}
    \State $\text{d\_z[desired\_idx]} \gets \text{temp\_z[temp\_idx]}$
    \State $\text{desired\_idx} \gets \text{desired\_idx} + \text{blockDim.x}$
    \State $\text{temp\_idx} \gets \text{temp\_idx} + \text{PATHS}$
\EndFor
\end{algorithmic}
\end{algorithm}

Shown in in Figure~\ref{fig:QuasiVariableTransformation} is the input and output ordering of random variables. We see that $B$ variables, where $B$ is BLOCK\_DIM, are taken from each dimension and placed next to each other. This process is repeated such that we have PATHS sets of random numbers from dimension $1$ to $d$. One set will be used by one block such that the BLOCK\_SIZE threads in that block simulate a single path each (one timestep uses one of the dimensions), with the random variable accesses being coalesced.

\begin{figure}[h]
    \centering
    \includegraphics[width=0.7\textwidth]{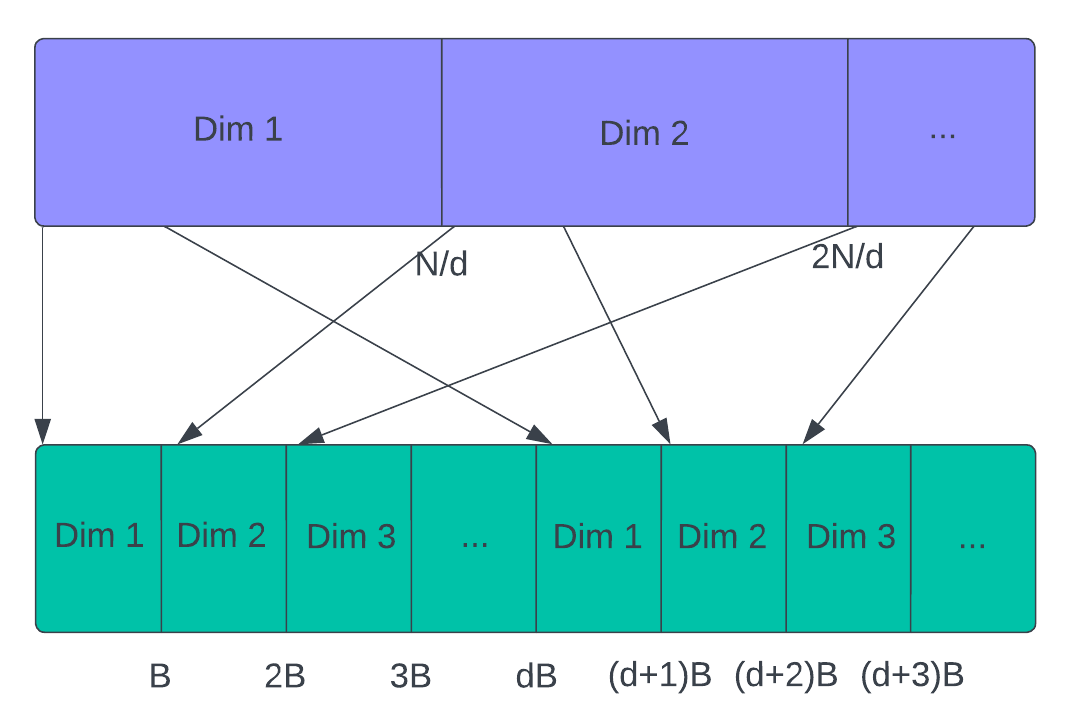}
    \caption{Transformation of cuRAND Sobol' numbers from input (top) to output (bottom) ordering.}
    \label{fig:QuasiVariableTransformation}
\end{figure}

In order to reduce the memory footprint and number of accesses, the results of path simulation are not stored for standard MC and standard QMC and required results are calculated on-the-fly during path simulation. This decision also reinforces the decision to encapsulate simulation inside of each product --- discussed in Section~\ref{sec:Products}. This allows us to store values required for calculation of the Greeks (such as $S_A$) whilst simulating the path, and use them in the later steps. The basic steps are outlined in Algorithm~\ref{alg:PathSimulation}. The separation of $S$ and $\widetilde{S}$ is necessary so that we are able to calculate the Greeks estimates as per section~\ref{sec:qmc-cpwmethod}. For QMC with Brownian bridge construction (see Section~\ref{sec:BBConstructionResults} for further description) we must store the intermediate Brownian bridge results to consume them for path generation afterwards.

\begin{algorithm}[hbt!]
\caption{Per-thread path simulation where $N$ is the number of simulated timesteps with $dt = 1/N$}\label{alg:PathSimulation}
\begin{algorithmic}[1]
\State $S \gets S_0$
\State $Z \gets \text{RandomNormals[ind]}$
\State $W_1 \gets sqrt(dt) * Z$
\State $\widetilde{W}_1 \gets W_1$
\For{$i \gets 1..N$}
    \State $\text{ind} \gets \text{ind} + \text{blockDim.x}$ \Comment{Coalesced reads, when blockDim.x is a multiple of 32}
    \State $Z \gets \text{RandomNormals[ind]}$
    \State $\widetilde{W}_i \gets \widetilde{W}_i + sqrt(dt) * Z$
    \State $\widetilde{S} \gets S_0 * \exp{(\omega * (n-1)*dt + \sigma * (\widetilde{W} - W_1))}$
    \State $S \gets \widetilde{S} * \exp{(\omega * dt + \sigma * W_1)}$
\EndFor
\end{algorithmic}
\end{algorithm}

\subsection{Products} \label{sec:Products}

It is required to calculate the prices and sensitivities of a variety of options and the functions to do so typically vary between different option types. However, the overall process is the same for pricing any derivative, namely: simulate paths of the underlying asset, followed by calculating the prices and Greeks given the simulated path. These two requirements are that of any option and as such we combine them into a \emph{product}. In this paper we focus on three types of exotic option: arithmetic Asian, binary Asian and lookback. For derivations of the Greek estimates as in Section~\ref{sec:qmc-cpwmethod} see Section~\ref{sec:GreeksCalculation}. Each of these products implements its own path simulation and Greeks calculation method.

Inheritance and virtual functions are widely-used in standard C++ and similar programming languages, however there are many more restrictions with CUDA. Due to having separate address spaces, copying objects with virtual functions from host memory to device memory can be tricky. To avoid unnecessary complexity, we avoid the use of inheritance directly in kernels (on device) and use them only to aid readability and development. To avoid inheritance directly, we make use of C++ templates. That is, kernels which are used for multiple option types are templatised, so that at compile time distinct versions of the kernel are generated for each option. From this we obtain the same benefits from inheritance such as minimal repetition of code, without having to copy objects with virtual function tables across address spaces or perform any casts.

Each thread instantiates its own local copy of the product which has member fields for values such as the underlying's price at the current timestep, running averages, and index to the current random variable. The \emph{SimulatePath} method is called and that thread performs a single simulation for the product, calculating any intermediate values such as the average underlying price or the inner sum of the vega estimate. The final call is to the \emph{CalculatePayoffs} function which calculates the price of the option and Greeks, then places these values back into the global struct of arrays of results.

\subsection{Antithetic variables}

As a variance reduction technique we have used antithetic variables. Using the already generated random normal variables for the standard MC simulation, we take their complement and simulate a second path from which another set of estimates are calculated. The estimates from the standard and antithetic paths can then be combined to produce the variance-reduced final estimate. Adding antithetic variables requires minimal storage on device as we only need to add fields to our products struct that represent the antithetic counterpart to the standard MC values such as $\widetilde{S}(t_j)$.

\subsection{Brownian bridge construction} \label{sec:BBConstructionResults}

For QMC, we have implemented Brownian bridge construction as a variance reduction method. As shown in Algorithm \ref{alg:PathSimulation} we generate the Brownian motion $\widetilde{W}_i$ from left to right (i.e. from $i=1\dots,d$). However, we may choose to generate the $\widetilde{W}_i$ in any order as long as we sample from the correct conditional distribution given the values already generated. Conditioning a Brownian motion on its endpoints produces a \emph{Brownian bridge}~\cite{glasserman2004monte}. The basic idea is that we generate the final value $\widetilde{W}_d$, then continue to fill in each intermediate value: $\widetilde{W}_{d/2}$, then $\widetilde{W}_{d/4}$ and $\widetilde{W}_{3d/4}$ etc, until all values are calculated. For further explanation of how the conditional mean and variance are derived, the reader is referred to Section~3.1 of~\cite{glasserman2004monte}.

Our implementation does not construct the path directly using a Brownian bridge, but rather uses the bridge to calculate the increments in the path. This allows us to construct $\widetilde{W}$ simply by iterating through the output of the Brownian bridge construction and adding it to the previous value. Algorithm~\ref{alg:BrownianBridgeConstruction} demonstrates the process of constructing the Brownian bridge increments.

\begin{algorithm}[hbt!]
\caption{Construction of Brownian bridge increments where the number of timesteps is equal to $2^m$. \emph{idx\_zero} is passed to each thread as the first index into the global path array.}\label{alg:BrownianBridgeConstruction}
\begin{algorithmic}[1]
\State $\text{path[idx\_zero]} \gets \text{d\_z[idx]}$ \Comment{Put first random variable (representing terminal value) in path}
\For{$k \gets 1..m$}
    \State $i \gets 2^k - 1$
    \For{$j \gets 2^{k-1}-1..0$}
        \State $\text{idx} \gets \text{idx} + \text{blockDim.x}$ \Comment{Access next random variable}
        \State $z = \text{d\_z[idx]}$
        \State $a \gets 0.5 * \text{path[idx\_zero} + j * \text{blockDim.x]}$
        \State $b \gets \sqrt{1 / 2^{k+1}}$
        \State $\text{path[idx\_zero} + i * \text{blockDim.x]} \gets a - b * z$
        \State $i \gets i - 1$
        \State $\text{path[idx\_zero} + i * \text{blockDim.x]} \gets a + b * z$
        \State $i \gets i - 1$
    \EndFor
\EndFor
\end{algorithmic}
\end{algorithm}

Brownian bridge construction gives finer control over the overall structure of the simulated path as opposed to the standard recursion technique: we use only one random variable to generate the terminal value and then continue to add more and more detail to the rest of the path. Furthermore, when using Sobol' sequences, the first random variables are particularly well distributed leading to the terminal values also being well distributed. This is due to the fact that the initial coordinates of a Sobol' sequence have superior uniformity to that of higher-indexed coordinates~\cite{glasserman2004monte}. As the terminal value is often more important than other values in the path this can lead to less error in the estimates produced by Brownian bridge construction with Sobol' sequences. An example of how the path is generated as more points are sampled can be seen in Figure~\ref{fig:BrownianBridgePlots}.

The main downside with performing Brownian bridge construction rather than the standard approach is that we need to store the generated path to later consume to simulate the stock price in the variable separated form as per Section~\ref{sec:VariableSeparationPathSimulation}. This means we not only use more global memory on device but will also have a slower kernel runtime due to the increase in memory accesses. However, with this trade-off we expect to achieve a much smaller error in our estimates.

\begin{figure}[h]
    \centering
    \includegraphics[width=0.8\textwidth]{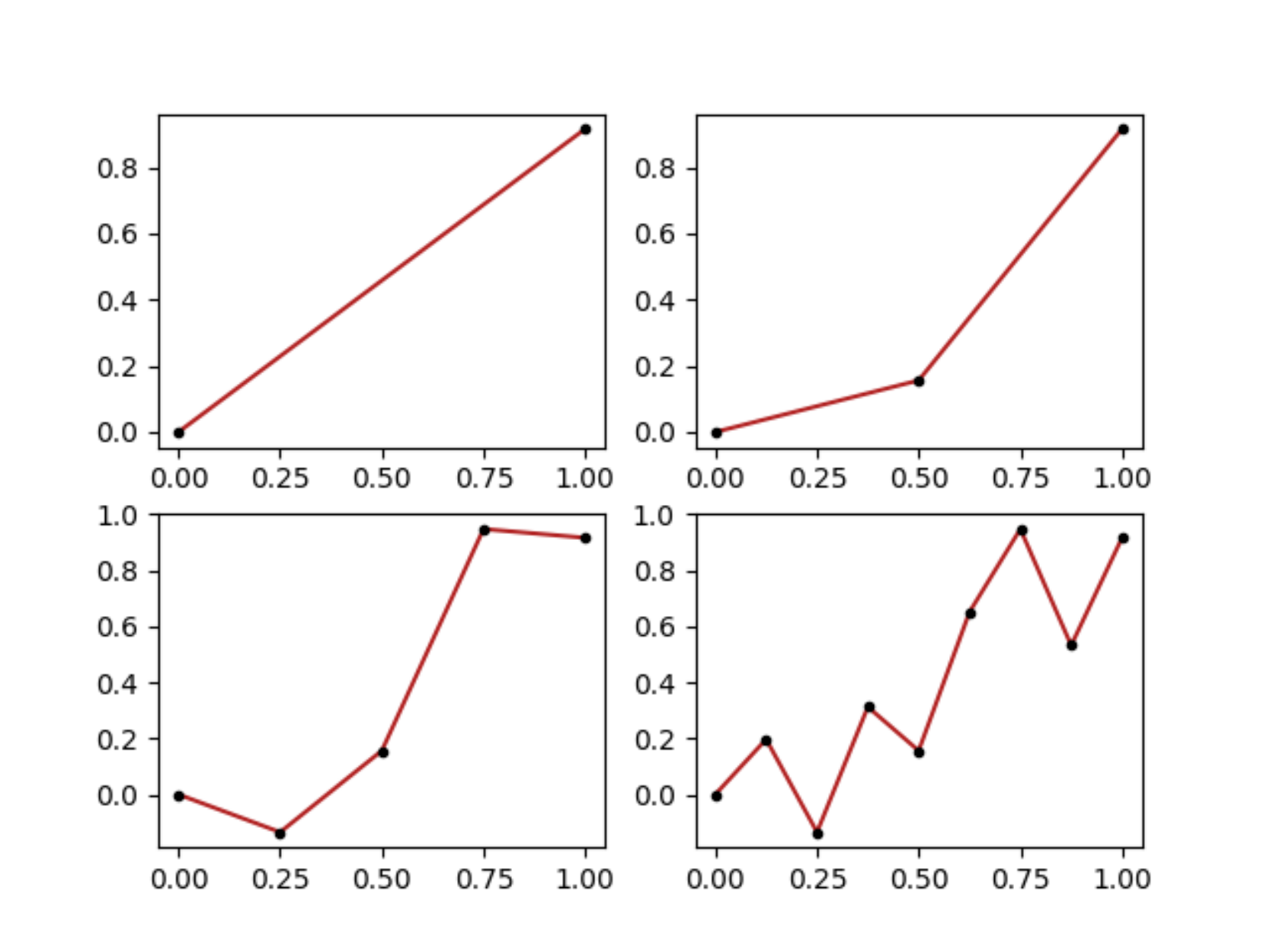}
    \caption{Brownian bridge construction after $1$, $2$, $4$ and $8$ points have been sampled conditional on the previous values generated.}
    \label{fig:BrownianBridgePlots}
\end{figure}

\subsection{Greeks calculation} \label{sec:GreeksCalculation}

The step of calculating Greeks is straightforward. Once we have simulated the path and saved the required values we simply need to evaluate the estimates and store them. Below we list the derived Greeks that are used to calculate estimates as per~\cite{ZhangConditionalQuasiMonteCarloMethod} following on from the example in Section~\ref{sec:BaDeltaExample}.

\subsubsection{Binary Asian Greeks}

As we have already shown the full derivation for the delta, we continue with the estimates for gamma and vega.

\begin{equation*}
\begin{aligned}
    \text{gamma: }\frac{{\partial}^2 G}{\partial {S(0)}^2} &= \frac{e^{-rT}}{{S(0)}^2\sigma \sqrt{t_1}}\phi(\psi_d)\left(\frac{\psi_d}{\sigma \sqrt{t_1}} - 1\right). \\[10pt]
\end{aligned}
\end{equation*}

\begin{equation*}
\begin{aligned}
    \text{vega: }\frac{{\partial}^2 G}{\partial \sigma} &= e^{-rT}\phi(\psi_d) \left[ \frac{1}{d\sigma \sqrt{t_1}\widetilde{S}_A} \sum_{j=1}^d{\widetilde{S}(t_j) (\widetilde{B}(t_j - t_1) - \sigma(t_j - t_1))} + \frac{\psi_d}{\sigma} - \sqrt{t_1} \right]. \\[10pt]
\end{aligned}
\end{equation*}

Note that the sum inside of the vega calculation is an example of one of the values that is calculated on-the-fly during the path simulation, allowing us to disregard storing the path for standard MC and QMC and storing single precision values only.

\subsubsection{Arithmetic Asian Greeks}

By taking the conditional expectation we obtain the smoothed payoff
\begin{equation*}
    G(\theta, \boldsymbol{z}) = e^{r(t_1 - T)} \widetilde{S}_A \left[1 - \Phi(\psi_d - \sigma \sqrt{t_1} \right] - e^{-rT}K\left[ 1 - \Phi(\psi_d) \right]
\end{equation*}

We can now differentiate with respect to our parameters of interest to obtain the following estimates.
\begin{equation*}
\begin{aligned}
    \text{delta: }\frac{\partial G}{\partial {S(0)}} &= e^{r(t_1 - T)} \frac{\widetilde{S}_A}{S(0)}\left[1 - \Phi(\psi_d - \sigma \sqrt{t_1}) \right]. \\[10pt]
\end{aligned}
\end{equation*}
\begin{equation*}
\begin{aligned}
    \text{gamma: }\frac{{\partial}^2 G}{\partial {S(0)}^2} &= \frac{Ke^{-rT}}{{S(0)}^2\sigma\sqrt{t_1}} \phi(\psi_d). \\[10pt]
\end{aligned}
\end{equation*}
\begin{equation*}
\begin{aligned}
    \text{vega: }\frac{{\partial} G}{\partial \sigma} &= e^{r(t_1 - T)} \left[1 - \Phi(\psi_d - \sigma \sqrt{t_1}) \right] \frac{1}{d}\sum_{j=1}^d{\widetilde{S}(t_j) (\widetilde{B}(t_j - t_1) - \sigma(t_j - t_1))} + Ke^{-rT}\phi(\psi_d)\sqrt{t_1}. \\[10pt]
\end{aligned}
\end{equation*}

\subsubsection{Lookback Greeks}

Again, we take the conditional expectation to obtain the smoothed payoff
\begin{equation*}
    G(\theta, \boldsymbol{z}) = e^{r(t_1 - T)} \widetilde{S}_{max} \left[1 - \Phi(\psi_d - \sigma \sqrt{t_1} \right] - e^{-rT}K\left[ 1 - \Phi(\psi_d) \right],
\end{equation*}
where $\widetilde{S}_{max}$ is the maximum value of $\widetilde{S}(t_j)$ for $j = 1,\dots,d$, and $\psi_d = (\ln{K} - \ln{\widetilde{S}_{max}} - \omega t_1) / \sigma \sqrt{t_1}$. By taking differentiation with respect to our parameters we obtain the estimates
\begin{equation*}
\begin{aligned}
    \text{delta: }\frac{\partial G}{\partial {S(0)}} &= e^{r(t_1 - T)} \frac{\widetilde{S}_{max}}{S(0)}\left[1 - \Phi(\psi_d - \sigma \sqrt{t_1}) \right]. \\[10pt]
\end{aligned}
\end{equation*}
\begin{equation*}
\begin{aligned}
    \text{gamma: }\frac{{\partial}^2 G}{\partial {S(0)}^2} &= \frac{Ke^{-rT}}{{S(0)}^2\sigma\sqrt{t_1}} \phi(\psi_d). \\[10pt]
\end{aligned}
\end{equation*}
\begin{equation*}
\begin{aligned}
    \text{vega: }\frac{{\partial} G}{\partial \sigma} &= e^{r(t_1 - T)} \left[1 - \Phi(\psi_d - \sigma \sqrt{t_1}) \right] \frac{1}{d}\sum_{j=1}^d{\widetilde{S}(t_j) (\widetilde{B}(t_j - t_1) - \sigma(t_j - t_1)) \boldsymbol{1}\{\widetilde{S}(t_j) = \widetilde{S}_{max}\}}\\ 
    &\quad + Ke^{-rT}\phi(\psi_d)\sqrt{t_1}. \\[10pt]
\end{aligned}
\end{equation*}

\subsection{Likelihood Ratio estimates}

As a baseline for the error in the Greek estimates, we implement the LR method through Monte Carlo simulation. Taking the ideas in Section~\ref{sec:LikelihoodRatioMethod} we apply LR to our set of options. The expression given in~(\ref{eqn:LRUnbiasedEstimate}) shows that
\begin{equation*}
    f(X) \frac{g^\prime_\theta(X)}{g_\theta(X)},
\end{equation*}
is an unbiased estimator of the derivative of $\mathbb{E}[Y]$ with respect to parameter $\theta$. The expression $g^\prime_\theta(X)/g_\theta(X)$ is commonly referred to as the \emph{score}. Calculating Greeks using LR simplifies to calculating the product of the discounted payoff and the relevant score for the Greek. 

Below are listed the scores for the Greeks of each of the three options we are concerned with.
\begin{equation*}
\begin{aligned}
    \text{delta: }\frac{Z_1}{S(0)\sigma\sqrt{t_1}}. \\[10pt]
\end{aligned}
\end{equation*}
\begin{equation*}
\begin{aligned}
    \text{gamma: }\frac{Z_1^2 - 1}{{S(0)}^2 \sigma^2 t_1} - \frac{Z_1}{{S(0)}^2 \sigma \sqrt{t_1}}. \\[10pt]
\end{aligned}
\end{equation*}
\begin{equation*}
\begin{aligned}
    \text{vega: }\sum_{j=1}^d{\frac{Z_j^2 - 1}{\sigma} - Z_j\sqrt{t_1}}. \\[10pt]
\end{aligned}
\end{equation*}

Note that the scores for the three options are equal and the difference between the estimates is simply the form of the payoff.

\subsection{CPU implementation}

To demonstrate the superior speed when using GPUs we implement a naive, sequential Monte Carlo simulation with the same form of estimates from the aforementioned sections. The implementation has the general form shown in Algorithm~\ref{alg:EstimateExpectedPayoff}. The random normal variables generated for use in the GPU simulation are reused by the CPU simulation, in which a single thread performs \emph{NPATH} simulations of $N$ timesteps each. After each path simulation the estimates are calculated and stored in the results struct in the same way that a single GPU thread does.

\section{Results} \label{sec:Results}

To demonstrate the effectiveness of the QMC-CPW method from Section~\ref{sec:qmc-cpwmethod} we run many simulations on the GPU and calculate the \emph{variance reduction factors~(VRFs)} for multiple methods. Using the Likelihood Ratio estimate as the baseline for variance, the VRF for a method is calculated as 
\begin{equation*}
    \frac{\sigma^2_0}{\sigma^2},
\end{equation*}
where $\sigma^2_0$ is the variance in the LR estimate for the Greek. For all methods the estimates for the Greeks are calculated over $P$ number of paths of $N$ timesteps, such that the estimate from a single path is given as
\begin{equation*}
    C^{(\ell)} = {F(\theta,\boldsymbol{z}_\ell}),
\end{equation*}
where $\boldsymbol{z}_\ell$ is a vector of $N$ normal random variables and $F(\theta, x)$ is the underlying function we wish to estimate (e.g. the delta estimate for an arithmetic Asian option). To calculate the error in the estimate we perform $L$ independent runs of the simulation with $P$ fixed such that the final estimate is given as
\begin{equation*}
    C = \frac{1}{L} \sum_{\ell=1}^{L} C_P^{(\ell)},
\end{equation*}
where $C_P^{(\ell)}$ is the estimate from the $\ell$th run over $P$ paths. Finally, the error in the estimate is calculated as follows:
\begin{equation*}
    \sigma = \sqrt{\frac{1}{L} \sum_{\ell=1}^{L} {(C - C_P^{(\ell)})}^2}.
\end{equation*}

For delta, gamma and vega estimation we compare four methods: standard Monte Carlo with CPW estimates (MC-CPW), Monte Carlo with antithetic variables and CPW estimates (MC+AV-CPW), Quasi-Monte Carlo with CPW estimates (QMC-CPW), and finally Quasi-Monte Carlo with Brownian bridge construction and CPW estimates (QMC+BB-CPW). Following a similar style as in~\cite{ZhangConditionalQuasiMonteCarloMethod} we perform the simulations over a range of strike prices $K=90,100,110$, and two values for the number of discrete time steps $d=64,256$. We denote the option as ``in the money'' at $K=90$, ``at the money'' at $K=100$ and ``out the money'' at $K=110$. The number of paths, initial stock price, volatility, and risk-free interest rate are all constant and equal for each option type with $P = 2^{15}$,  $S(0) = 100$, $\sigma = 0.2$, and $r = 0.1$. The expiration date for each option $T = 1.0$, or one year. We perform $L=500$ independent runs for all methods. The VRFs for arithmetic, binary and lookback options are presented in Tables~\ref{tbl:vrfs-arithmetic}--\ref{tbl:vrfs-lookback} respectively. Later we discuss the behaviour of the error in Greek estimates as we increase the number of path simulations per independent run. Information about the \emph{Tesla T4} GPU and the specifications of the CUDA toolkit that was used to collect the results can be found in Appendix~\ref{app:gpu-cuda-specs}.

We can make the following observations from the experimental results:

\begin{itemize}
    \item The QMC+BB-CPW method is the most accurate in almost all cases. This is due to the combination of the CPW method which smooths the integrand, allowing for QMC method to work more efficiently, and the Brownian bridge construction which further reduces variance through the methods described in Section~\ref{sec:BBConstructionResults}.
    \item For the arithmetic Asian option we see QMC+BB-CPW as the best method in all experiments, with VRFs in the hundreds of thousands, and in many cases more than $10$x accurate in comparison to QMC-CPW and MC+AV-CPW. When looking at the VRFs for gamma estimates of the arithmetic Asian option (Table~\ref{tbl:vrfs-arithmetic}), MC+AV-CPW outperforms QMC-CPW and this could be due to MV+AV-CPW effectively simulating twice as many paths (standard + antithetic paths) which of course helps to reduce the variance. However, this is not the case for the delta and vega estimates which is interesting to note.
    \item Strike price does affect the performance of many experiments, particularly for the delta and gamma estimates, in which we see an increase in the strike leading to a decrease in VRF.
    \item We discuss dimensionality later, but it also has an effect on the accuracy and becomes more apparent for QMC methods.
\end{itemize}

\newcolumntype{L}{>{$}c<{$}}

\begin{table}[h]
\centering
\begin{tabular}{ L L L L L L L L } 
 \toprule
 \text{Greeks} & K & d & \text{LR+MC} & \text{MC-CPW} & \text{MC+AV-CPW} & \text{QMC-CPW} & \text{QMC+BB-CPW} \\
 \midrule
 \text{delta} & 90  & 64  & 1 & 623     & 3{,}209 & 5,784    & \boldsymbol{154{,}860} \\ 
              &     & 256 & 1 & 2{,}159 & 9{,}527 & 11{,}976 & \boldsymbol{106{,}806} \\
              & 100 & 64  & 1 & 106     & 963     & 903      & \boldsymbol{52{,}689} \\
              &     & 256 & 1 & 353     & 2{,}702 & 1{,}735  & \boldsymbol{34{,}478} \\
              & 110 & 64  & 1 & 35      & 172     & 207      & \boldsymbol{13{,}226} \\
              &     & 256 & 1 & 103     & 423     & 445      & \boldsymbol{7{,}645} \\
 \\
 \text{vega} & 90  & 64  & 1 & 471     & 1{,}566 & 14,603   & \boldsymbol{442{,}513} \\ 
             &     & 256 & 1 & 1{,}595 & 5{,}424 & 30{,}894 & \boldsymbol{340{,}858} \\
             & 100 & 64  & 1 & 294     & 759     & 7{,}770  & \boldsymbol{376{,}285} \\
             &     & 256 & 1 & 967     & 2{,}540 & 18{,}162 & \boldsymbol{633{,}051} \\
             & 110 & 64  & 1 & 113     & 289     & 3{,}195  & \boldsymbol{119{,}816} \\
             &     & 256 & 1 & 330     & 917     & 6{,}701  & \boldsymbol{294{,}813} \\
 \\
 \text{gamma} & 90  & 64  & 1 & 20{,}393  & 49{,}141  & 28{,}067  & \boldsymbol{271{,}351} \\ 
              &     & 256 & 1 & 108{,}667 & 275{,}370 & 134{,}644 & \boldsymbol{487{,}940} \\
              & 100 & 64  & 1 & 3{,}814   & 9{,}433   & 5{,}427   & \boldsymbol{75{,}020} \\
              &     & 256 & 1 & 20{,}967  & 49{,}834  & 21{,}116  & \boldsymbol{72{,}558} \\
              & 110 & 64  & 1 & 1{,}101   & 2{,}468   & 1{,}477   & \boldsymbol{23{,}085} \\
              &     & 256 & 1 & 5{,}977   & 13{,}770  & 6{,}147   & \boldsymbol{25{,}695} \\
 \bottomrule
\end{tabular}
\caption{VRFs for arithmetic Asian option on GPU with $2^{15}$ paths. $S(0) = 100$, $\sigma = 0.2$, $r = 0.1$ and $T = 1$.}
\label{tbl:vrfs-arithmetic}
\end{table}

\begin{itemize}
    \item For delta and vega estimates of the binary Asian option (Table \ref{tbl:vrfs-binary}) we see QMC+BB-CPW outperforming all other methods and taking advantage of the increased smoothness of the integrand.
    \item We see little or no improvement of QMC-CPW over MC-CPW for all estimates of the binary option which could be an indication of the limitations of QMC in high dimensions.
    \item We also see this in the gamma estimates for the binary Asian option, where even QMC+BB-CPW is outperformed by MC+AV-CPW for all of the experiments with $256$ timesteps. A technique to reduce the effective dimension of the problem such as Principle Component Analysis~(PCA) would likely remove these differences and result in a substantial decrease in error for the QMC methods.
    \item The binary Asian option results in some of the smallest VRFs for all Greek estimates especially for the delta and gamma.
    \item Again, we see the strike price having a large impact on the VRFs. For example, the delta estimate with $K=90$ over $256$ timesteps in Table \ref{tbl:vrfs-binary} is $714$ and decreases to $150$ for $K=110$.
\end{itemize}

\begin{table}[h]
\centering
\begin{tabular}{ L L L L L L L L } 
 \toprule
 \text{Greeks} & K & d & \text{LR+MC} & \text{MC-CPW} & \text{MC+AV-CPW} & \text{QMC-CPW} & \text{QMC+BB-CPW} \\
 \midrule
 \text{delta} & 90  & 64  & 1 & 109 & 247 & 150 & \boldsymbol{1{,}447} \\ 
              &     & 256 & 1 & 159 & 389 & 197 & \boldsymbol{714} \\
              & 100 & 64  & 1 & 43  & 123 & 58  & \boldsymbol{830} \\
              &     & 256 & 1 & 64  & 150 & 64  & \boldsymbol{221} \\
              & 110 & 64  & 1 & 23  & 69  & 32  & \boldsymbol{497} \\
              &     & 256 & 1 & 35  & 89  & 36  & \boldsymbol{150} \\
 \\
 \text{vega} & 90  & 64  & 1 & 326     & 733     & 447     & \boldsymbol{4{,}227} \\ 
             &     & 256 & 1 & 481     & 1{,}168 & 593     & \boldsymbol{2{,}114} \\
             & 100 & 64  & 1 & 771     & 2{,}078 & 1{,}176 & \boldsymbol{12{,}571} \\
             &     & 256 & 1 & 1{,}419 & 3{,}405 & 1{,}502 & \boldsymbol{5{,}111} \\
             & 110 & 64  & 1 & 839     & 1{,}965 & 1{,}167 & \boldsymbol{9{,}232} \\
             &     & 256 & 1 & 1{,}976 & 4{,}617 & 1{,}950 & \boldsymbol{7{,}803} \\
 \\
 \text{gamma} & 90  & 64  & 1 & 363 & 713                  & 376 & \boldsymbol{999} \\ 
              &     & 256 & 1 & 691 & \boldsymbol{1{,}446} & 673 & 883 \\
              & 100 & 64  & 1 & 136 & 201                  & 126 & \boldsymbol{784} \\
              &     & 256 & 1 & 201 & \boldsymbol{415}     & 212 & 392 \\
              & 110 & 64  & 1 & 79  & 137                  & 67  & \boldsymbol{355} \\
              &     & 256 & 1 & 116 & \boldsymbol{237}     & 117 & 179 \\
 \bottomrule
\end{tabular}
\caption{VRFs for binary Asian option on GPU with $2^{15}$ paths. $S(0) = 100$, $\sigma = 0.2$, $r = 0.1$ and $T = 1$.}
\label{tbl:vrfs-binary}
\end{table}

\begin{table}[h]
\centering
\begin{tabular}{ L L L L L L L L } 
 \toprule
 \text{Greeks} & K & d & \text{LR+MC} & \text{MC-CPW} & \text{MC+AV-CPW} & \text{QMC-CPW} & \text{QMC+BB-CPW} \\
 \midrule
 \text{delta} & 90  & 64  & 1 & 7{,}020  & 58{,}906  & 382{,}145     & \boldsymbol{2{,}631{,}721} \\ 
              &     & 256 & 1 & 26{,}665 & 187{,}898 & 1{,}135{,}848 & \boldsymbol{7{,}737{,}083} \\
              & 100 & 64  & 1 & 1{,}635  & 12{,}183  & 21{,}857      & \boldsymbol{40{,}682} \\
              &     & 256 & 1 & 8{,}323  & 58{,}180  & 79{,}683      & \boldsymbol{171{,}880} \\
              & 110 & 64  & 1 & 233      & 1{,}899   & 1{,}896       & \boldsymbol{13{,}181} \\
              &     & 256 & 1 & 920      & 5{,}594   & 4{,}264       & \boldsymbol{23{,}354} \\
 \\
 \text{vega} & 90  & 64  & 1 & 501     & 2{,}333 & 10{,}667 & \boldsymbol{51{,}816} \\ 
             &     & 256 & 1 & 1{,}855 & 7{,}492 & 33{,}043 & \boldsymbol{165{,}179} \\
             & 100 & 64  & 1 & 311     & 1{,}420 & 6{,}580  & \boldsymbol{35{,}370} \\
             &     & 256 & 1 & 1{,}138 & 4{,}569 & 20{,}418 & \boldsymbol{103{,}536} \\
             & 110 & 64  & 1 & 178     & 870     & 4{,}601  & \boldsymbol{26{,}065} \\
             &     & 256 & 1 & 657     & 2{,}876 & 13{,}739 & \boldsymbol{69{,}210} \\
 \\
 \text{gamma} & 90  & 64  & 1 & 55{,}102{,}633 & 113{,}383{,}607 & 113{,}193{,}362 & \boldsymbol{129{,}220{,}281} \\ 
              &     & 256 & 1 & 1.2 \times 10^{17}  & 4.2 \times 10^{17}          & 6.2 \times 10^{16}   & \boldsymbol{1.0 \times 10^{18}} \\
              & 100 & 64  & 1 & 27{,}235       & 72{,}792               & 89{,}333        & \boldsymbol{212{,}928} \\
              &     & 256 & 1 & 175{,}073      & 393{,}763              & 434{,}423       & \boldsymbol{604{,}285} \\
              & 110 & 64  & 1 & 9{,}787        & 24{,}450               & 13{,}755        & \boldsymbol{42{,}199} \\
              &     & 256 & 1 & 51{,}687       & \boldsymbol{123{,}922} & 60{,}037        & 112{,}398 \\
 \bottomrule
\end{tabular}
\caption{VRFs for lookback option on GPU with $2^{15}$ paths. $S(0) = 100$, $\sigma = 0.2$, $r = 0.1$ and $T = 1$.}
\label{tbl:vrfs-lookback}
\end{table}

\begin{itemize}
    \item For the lookback option (Table~\ref{tbl:vrfs-lookback}), we see some of the largest VRFs, particularly those for the gamma estimates.
    \item We also see just how great of an effect the strike price has on the lookback option: when $K=90$ and the option is in the money we can see a VRF of $1.0 \times 10^{18}$, whereas when the option is at the money and out the money we see estimates in the range of hundreds of thousands.
    \item For the delta and vega estimates QMC-CPW outperforms MC+AV-CPW for almost all experiments, except when $K=110$ for the delta estimate.
\end{itemize}

We also present graphs of the error in Greek estimates over a range of paths. The graphs in Figures~\ref{fig:ArithmeticPathErrorsK90}--\ref{fig:LookbackPathErrorsK110} are all calculated over $L=500$ independent runs with $P=2^i$ paths for $i \in [12,19]$, with $256$ timesteps each. The graphs for paths of $64$ timesteps are not included but we see similar behaviour to the graphs presented, and note that the earlier observations about dimensionality for the gamma estimates in Table~\ref{tbl:vrfs-binary} are maintained. We note the following observations:

\begin{itemize}
    \item QMC+BB-CPW tends to outperform other methods across all numbers of paths.
    \item Its advantage in gamma estimates typically appears to be much smaller except that of the arithmetic Asian option.
    \item For the delta and gamma estimates in Figure~\ref{fig:ArithmeticPathErrorsK90} we see QMC-CPW having little or no advantage over MC+AV-CPW.
    \item Vega estimates are typically the least accurate Greek.
    \item As the number of paths approaches $2^{19}$ we begin to see QMC+BB-CPW outperform all other methods for every Greek estimate.
\end{itemize}

\begin{figure}[H]
    \centering
    \includegraphics[width=1\textwidth]{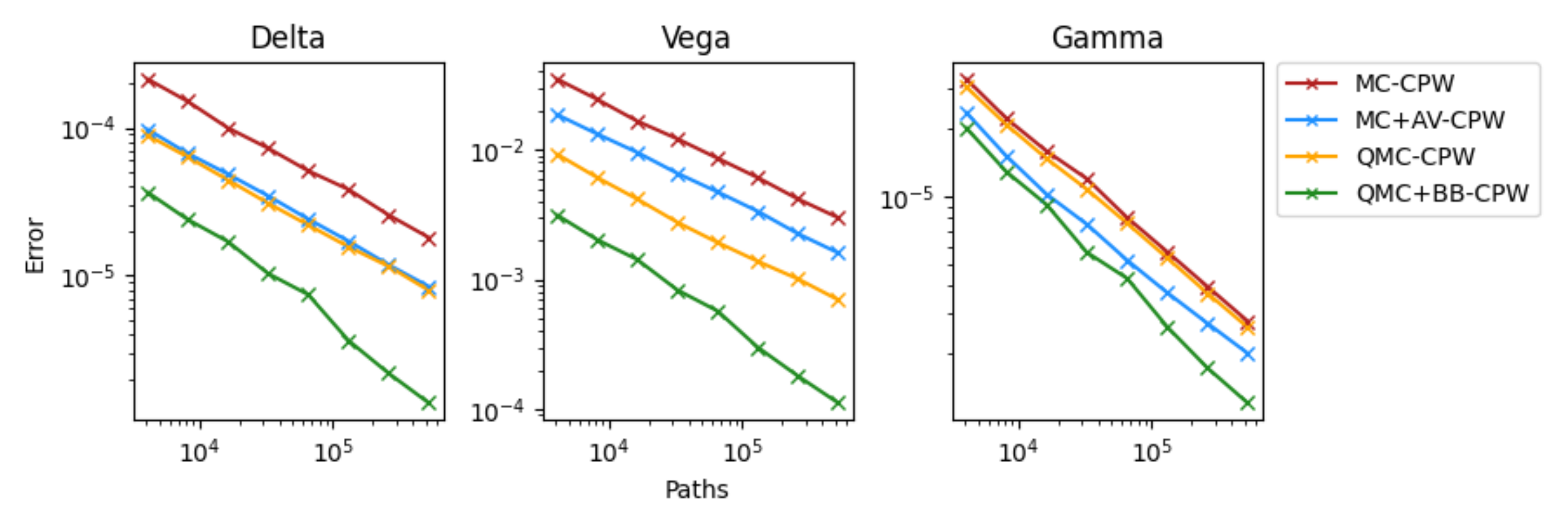}
    \caption{Errors in Greek estimates of an arithmetic Asian option with $S(0)=100$, $K=90$, $\sigma = 0.2$, $r=0.1$, $N=256$, and $T=1$ over $2^{12}$ to $2^{19}$ paths.}
    \label{fig:ArithmeticPathErrorsK90}
\end{figure}

\begin{figure}[H]
    \centering
    \includegraphics[width=1\textwidth]{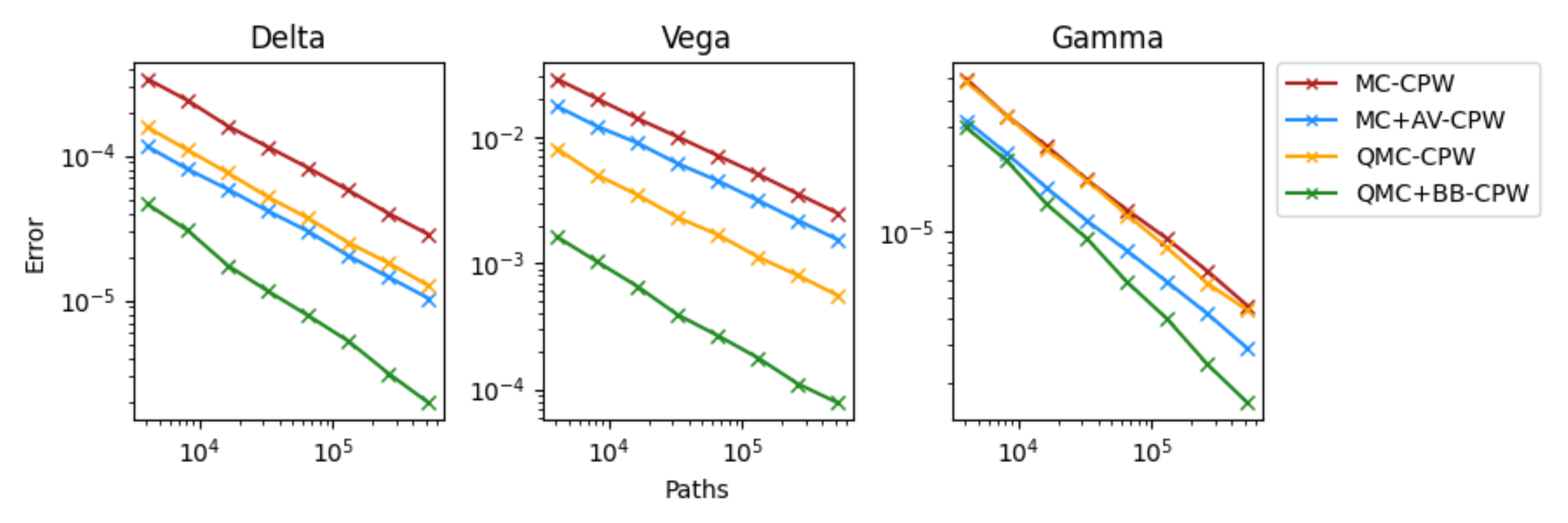}
    \caption{Errors in Greek estimates of an arithmetic Asian option with $S(0)=100$, $K=100$, $\sigma = 0.2$, $r=0.1$, $N=256$, and $T=1$ over $2^{12}$ to $2^{19}$ paths.}
    \label{fig:ArithmeticPathErrorsK100}
\end{figure}

\begin{figure}[H]
    \centering
    \includegraphics[width=1\textwidth]{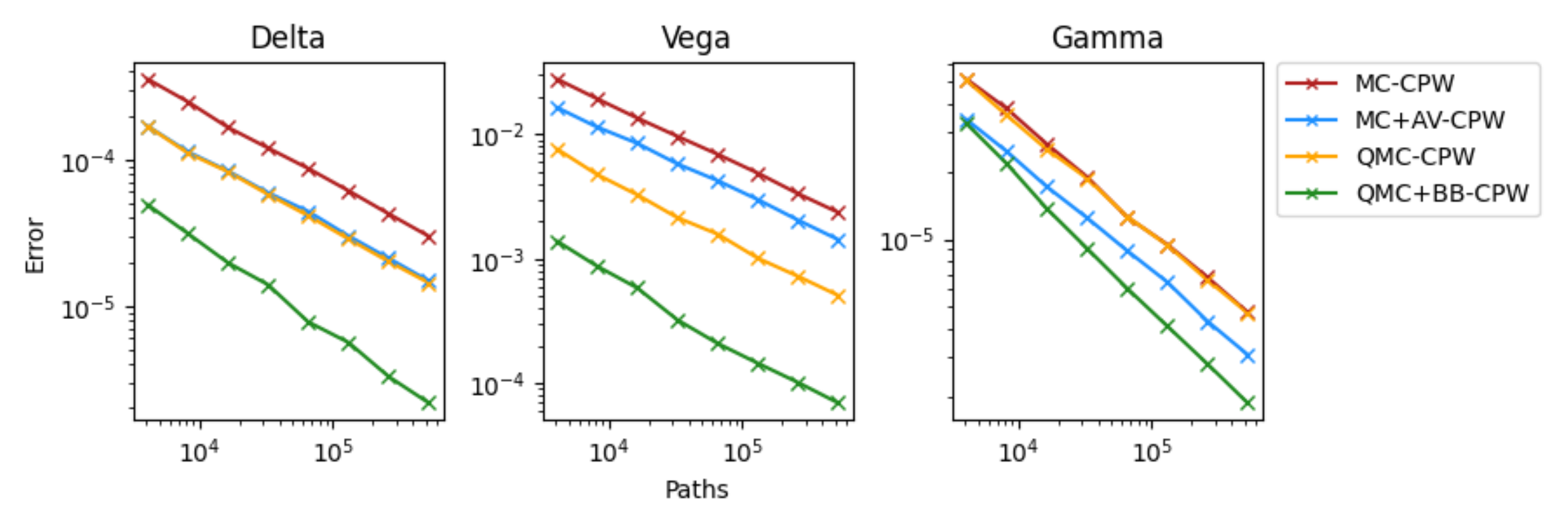}
    \caption{Errors in Greek estimates of an arithmetic Asian option with $S(0)=100$, $K=110$, $\sigma = 0.2$, $r=0.1$, $N=256$, and $T=1$ over $2^{12}$ to $2^{19}$ paths.}
    \label{fig:ArithmeticPathErrorsK110}
\end{figure}

\begin{itemize}
    \item For the arithmetic Asian estimates (Figures~\ref{fig:ArithmeticPathErrorsK90}--\ref{fig:ArithmeticPathErrorsK110}), QMC+BB-CPW is the best performing method across all number of paths and Greeks.
    \item For gamma estimates in figures \ref{fig:ArithmeticPathErrorsK90}--\ref{fig:ArithmeticPathErrorsK110} the advantage appears to increase as the number of paths increase.
    \item QMC-CPW is greatly outperformed by MC+AV-CPW for the arithmetic option's (Figures~\ref{fig:ArithmeticPathErrorsK90}--\ref{fig:ArithmeticPathErrorsK110}) gamma estimates of the arithmetic option whilst they perform similarly for delta.
    \item For the first order Greeks (delta and vega in Figures~\ref{fig:ArithmeticPathErrorsK90}--\ref{fig:ArithmeticPathErrorsK110}) QMC+BB-CPW has a large advantage over the other methods even at a small number of paths. However, for the second order Greek of gamma its error is roughly equal to that of MC+AV-CPW at a small number of paths and it only gains a noticeable advantage as the number of paths increases.
\end{itemize}

\begin{figure}[H]
    \centering
    \includegraphics[width=1\textwidth]{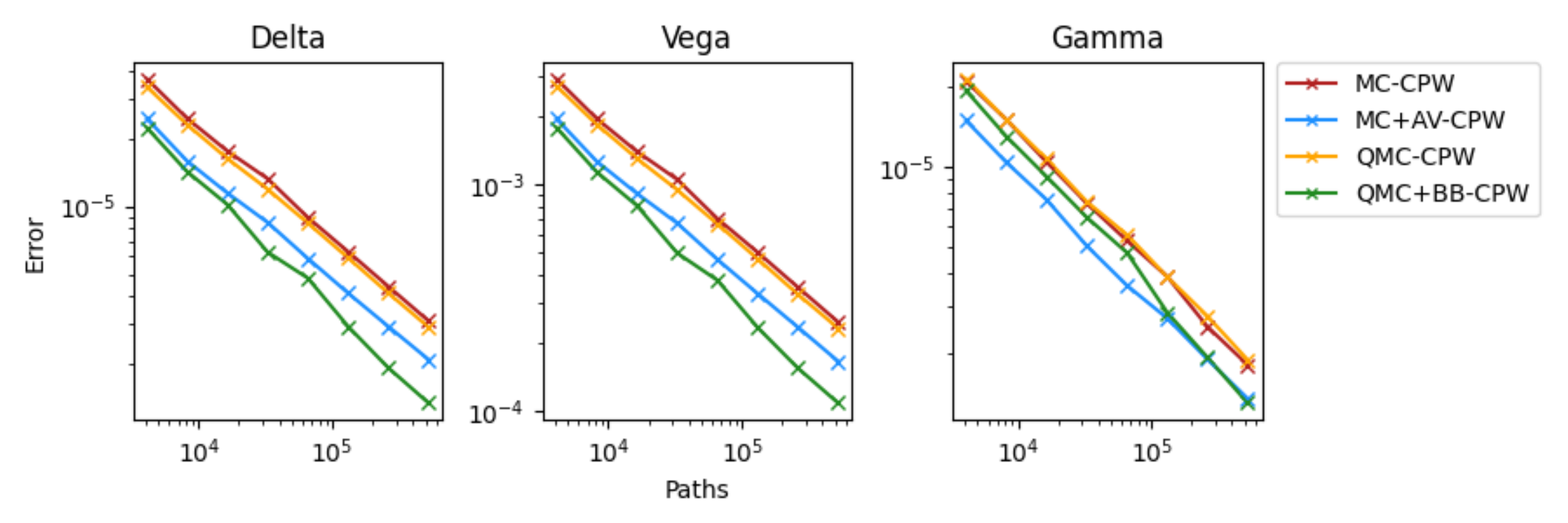}
    \caption{Errors in Greek estimates of a binary Asian option with $S(0)=100$, $K=90$, $\sigma = 0.2$, $r=0.1$, $N=256$, and $T=1$ over $2^{12}$ to $2^{19}$ paths.}
    \label{fig:BinaryPathErrorsK90}
\end{figure}

\begin{figure}[H]
    \centering
    \includegraphics[width=1\textwidth]{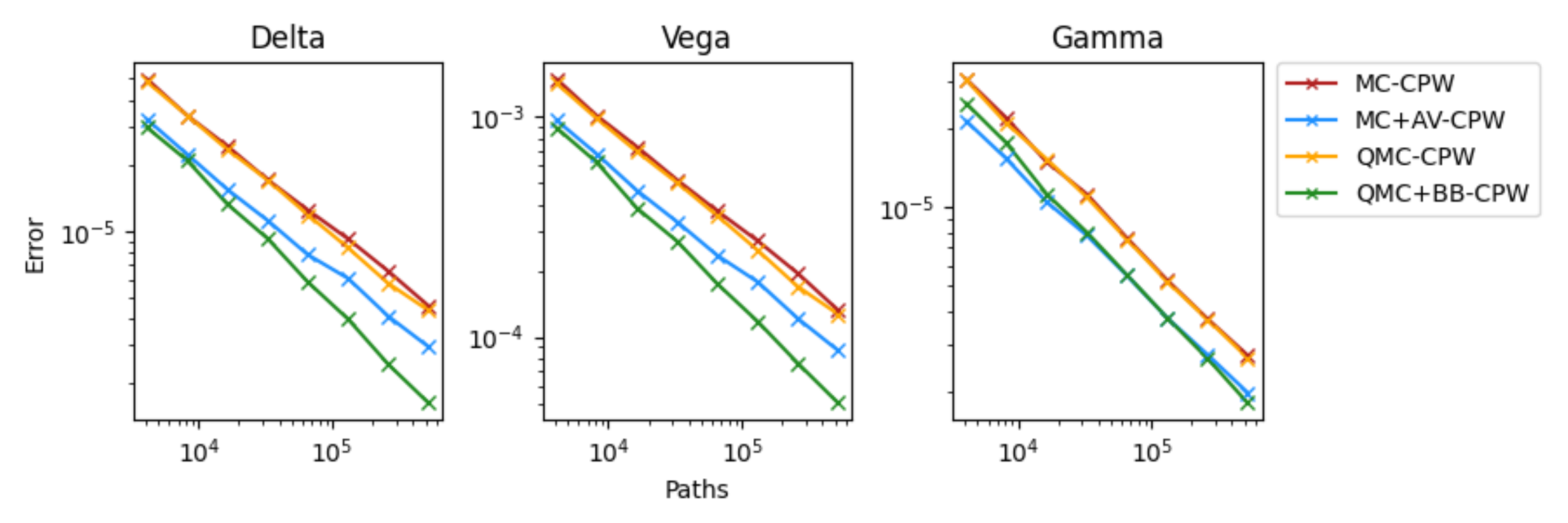}
    \caption{Errors in Greek estimates of a binary Asian option with $S(0)=100$, $K=100$, $\sigma = 0.2$, $r=0.1$, $N=256$, and $T=1$ over $2^{12}$ to $2^{19}$ paths.}
    \label{fig:BinaryPathErrorsK100}
\end{figure}

\begin{figure}[H]
    \centering
    \includegraphics[width=1\textwidth]{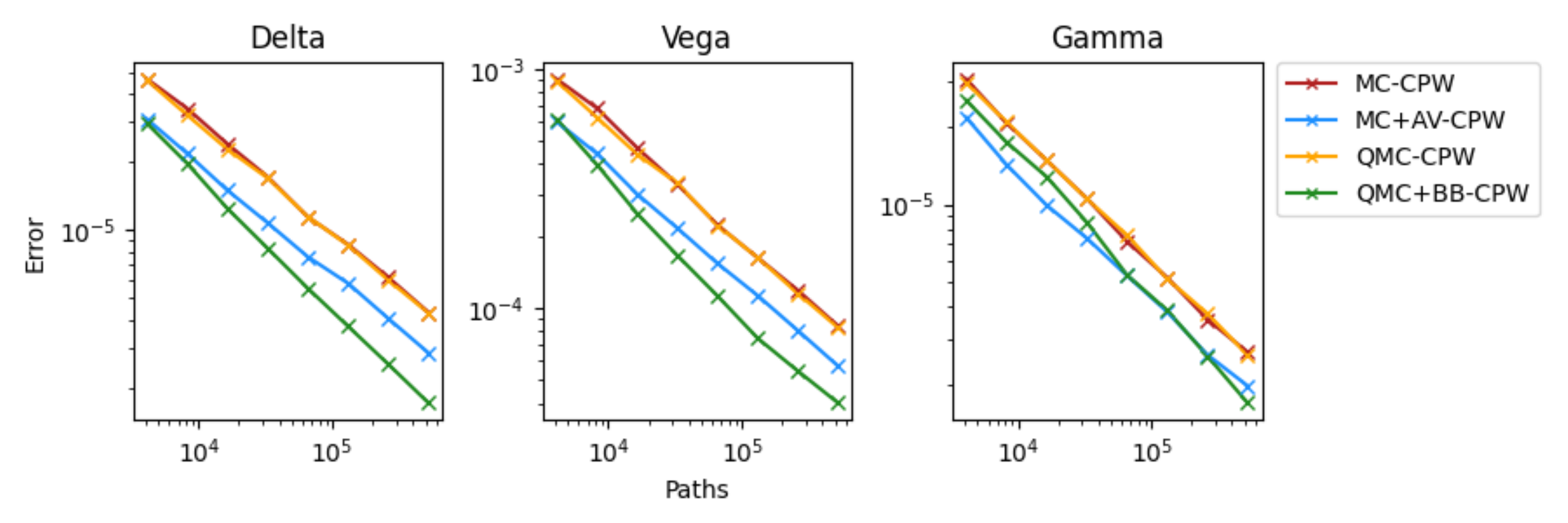}
    \caption{Errors in Greek estimates of a binary Asian option with $S(0)=100$, $K=110$, $\sigma = 0.2$, $r=0.1$, $N=256$, and $T=1$ over $2^{12}$ to $2^{19}$ paths.}
    \label{fig:BinaryPathErrorsK110}
\end{figure}

\begin{itemize}
    \item The errors in the estimates for the binary Asian option (figures \ref{fig:BinaryPathErrorsK90}-\ref{fig:BinaryPathErrorsK110}) are much closer than that of the arithmetic Asian.
    \item For delta and vega of the binary option in figures \ref{fig:BinaryPathErrorsK90}-\ref{fig:BinaryPathErrorsK110}, QMC+BB-CPW is the superior method across all number paths.
    \item MC+AV-CPW tends to match and often outperform QMC+BB-CPW when the number of paths is smaller for the binary option (figures \ref{fig:BinaryPathErrorsK90}-\ref{fig:BinaryPathErrorsK110}). In fact, we only see MC+AV-CPW outperformed for gamma at a very high path number ($2^{19}$).
    \item For all estimates of the binary option (figures \ref{fig:BinaryPathErrorsK90}-\ref{fig:BinaryPathErrorsK110}) we see almost no improvement with QMC-CPW over MC-CPW.
\end{itemize}

\begin{figure}[H]
    \centering
    \includegraphics[width=1\textwidth]{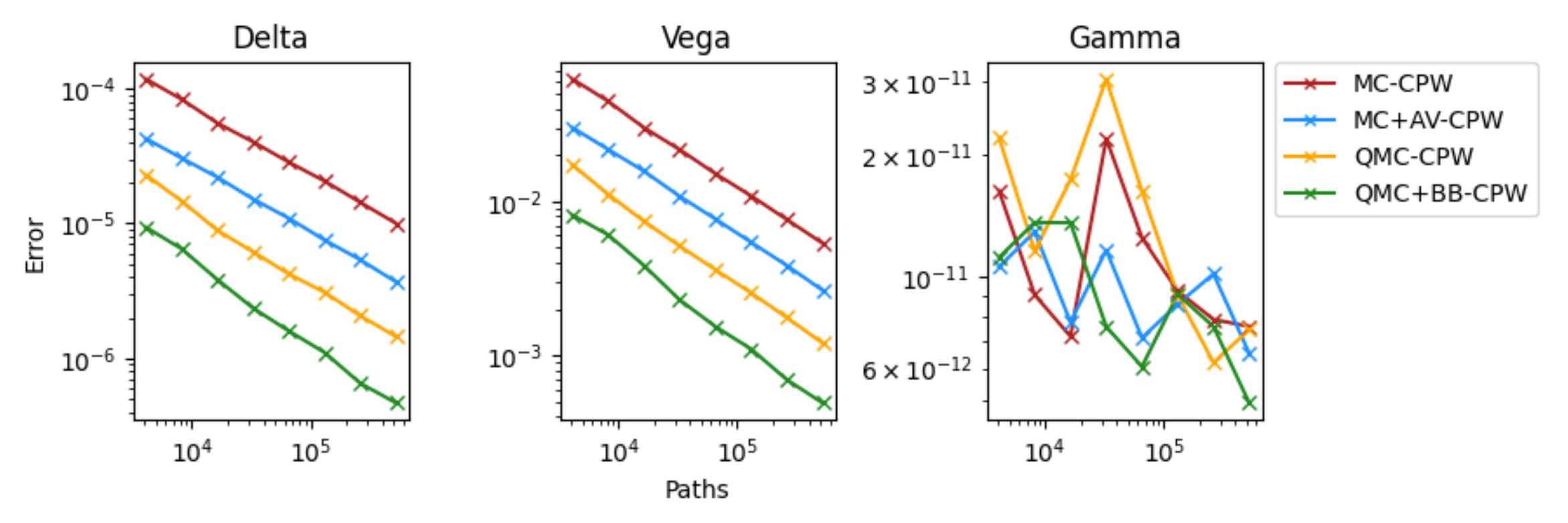}
    \caption{Errors in Greek estimates of a lookback option with $S(0)=100$, $K=90$, $\sigma = 0.2$, $r=0.1$, $N=256$, and $T=1$ over $2^{12}$ to $2^{19}$ paths.}
    \label{fig:LookbackPathErrorsK90}
\end{figure}

\begin{figure}[H]
    \centering
    \includegraphics[width=1\textwidth]{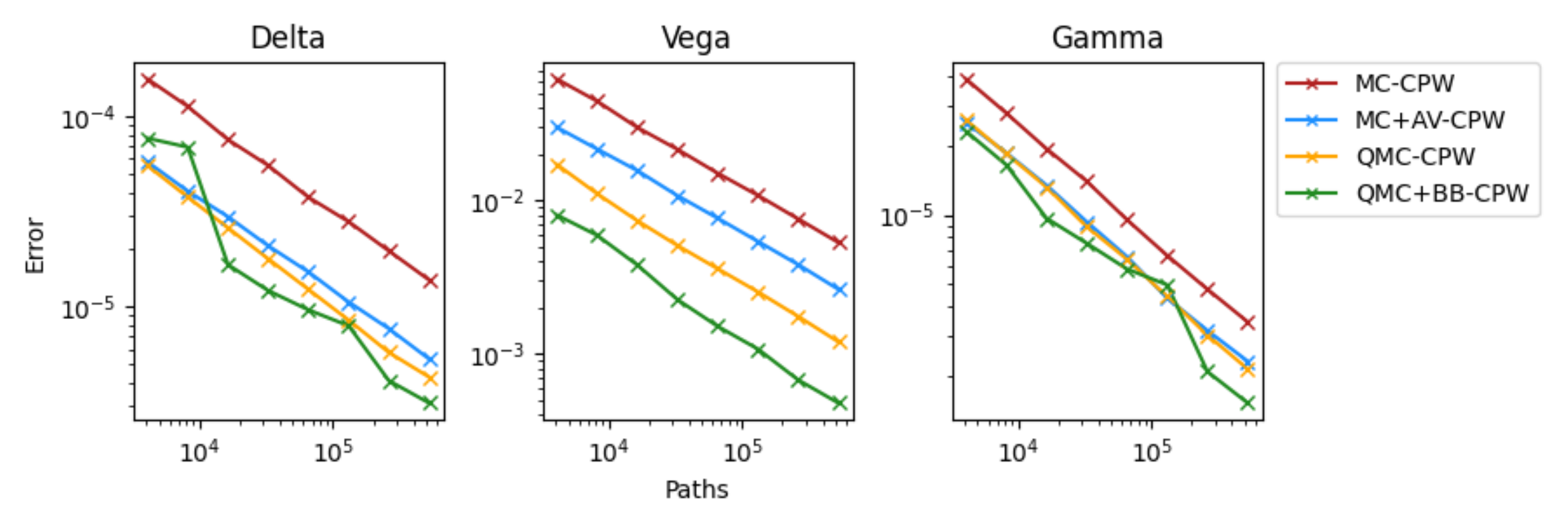}
    \caption{Errors in Greek estimates of a lookback option with $S(0)=100$, $K=100$, $\sigma = 0.2$, $r=0.1$, $N=256$, and $T=1$ over $2^{12}$ to $2^{19}$ paths.}
    \label{fig:LookbackPathErrorsK100}
\end{figure}

\begin{figure}[H]
    \centering
    \includegraphics[width=1\textwidth]{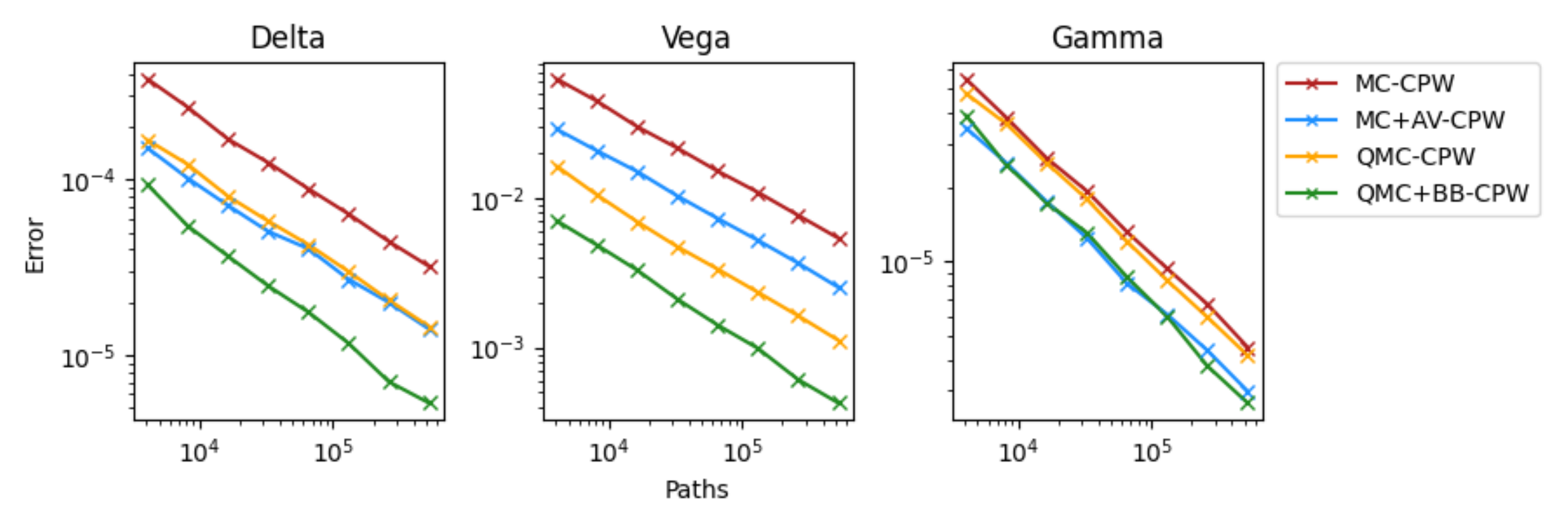}
    \caption{Errors in Greek estimates of a lookback option with $S(0)=100$, $K=110$, $\sigma = 0.2$, $r=0.1$, $N=256$, and $T=1$ over $2^{12}$ to $2^{19}$ paths.}
    \label{fig:LookbackPathErrorsK110}
\end{figure}

\begin{itemize}
    \item We see the largest variation in performance across the lookback estimates (Figures~\ref{fig:LookbackPathErrorsK90}--\ref{fig:LookbackPathErrorsK110}).
    \item For the gamma estimates when $K=90$ (in the money, figure \ref{fig:LookbackPathErrorsK90}) the errors are extremely small and do not follow the same monotonically decreasing trend we see in most other graphs.
    \item The vega estimates in figures~\ref{fig:LookbackPathErrorsK90}--\ref{fig:LookbackPathErrorsK110} are the most consistent where we see MC-CPW, MC+AV-CPW, QMC-CPW, QMC+BB-CPW as the order from largest error to smallest for $K=90,100,110$.
    \item When the option is at the money in Figure~\ref{fig:LookbackPathErrorsK100}, we see the least improvement of QMC+BB-CPW over QMC-CPW when compared to other options and estimates, where it is outperformed at a smaller number of paths and even matched at $2^{17}$ paths.
\end{itemize}

The final objective was to achieve a significant speed up over a CPU implementation. For the three methods where we do not store the path, we see speedups for a single kernel run when compared to the naive sequential CPU implementation upwards of $500$x for those experiments with $64$ timesteps per path, and upwards of $900$x for those with $256$ timesteps.

The overhead of accessing global memory on device becomes apparent when we see the difference in the speedup between the QMC+BB-CPW experiments and the other methods. Due to having to store the Brownian bridge path construction and then repeatedly accessing the array in global memory we see a significant decrease in speedups from the previously mentioned values to around $200$x for both $64$ and $256$ timesteps. It is interesting to note that the lookback option sees the greatest speed improvement over the CPU.

Although Tables~\ref{tbl:vrfs-arithmetic}--\ref{tbl:vrfs-lookback} show that MC+AV-CPW outperforms QMC+BB-CPW in some instances, from Figures~\ref{fig:ArithmeticPathErrorsK90}--\ref{fig:LookbackPathErrorsK110} we see that as the number of paths increases past $2^{15}$, which is the value used for the tables, QMC+BB-CPW tends to become the best performing method.
\section{Evaluation} \label{sec:Evaluation}

In this section the performance of our implementation is considered, in terms of VRFs and speedups, when compared with other similar solutions.

\subsection{Performance}

When compared with the implementation by Zhang and Wang~\cite{ZhangConditionalQuasiMonteCarloMethod} our most accurate method does not show as large VRFs as theirs. For example, many of their arithmetic Asian delta estimates (ours are in Table~\ref{tbl:vrfs-arithmetic}) are in the hundreds of millions whilst ours are in the hundreds of thousands. The implementation difference between our work and~\cite{ZhangConditionalQuasiMonteCarloMethod} is the variance reduction technique used with QMC. We use the Brownian bridge path construction, whereas Zhang and Wang use Gradient Principle Component Analysis~(GPCA)~\cite{gpca}. Whereas Brownian bridge is most effective for options whose terminal price is considered the most important value (e.g. European options), Asian options do not receive as great a variance reduction due to the form of their payoff. GPCA and PCA has been shown to reduce the effective dimension which makes QMC methods far more efficient, thus Zhang and Wang's implementation sees much better VRFs.

We can directly see the improvement of our implementation over that of the traditional pathwise and likelihood ratio methods simply from Tables~\ref{tbl:vrfs-arithmetic}--\ref{tbl:vrfs-lookback}. Noting the significantly better VRFs of QMC+BB-CPW in Section~\ref{sec:Results}, financial institutions would achieve much greater accuracy through the use of our implementation. Given how important calculating Greeks is for these institutions, the benefits from using our implementation are far and wide: a more precise understanding of individual products behaviour to input parameters can allow for a far better understanding of the overall risk a company has to the market. This allows a company to perhaps take on larger positions with more confidence in their exposure and give them the ability to better react to market events. In a more specific situation, having more accurate estimates for Greeks leads to better pricing of products, which can give a market participant an advantage over competitors.

As noted in Section~\ref{sec:Results}, as the number of paths increases, the error in the estimates from QMC+BB-CPW becomes the smallest of all the methods. We are able to comfortably simulate $2^{19}$ paths on the Tesla T4 GPU, thus the ever-present trade off between speed and accuracy is the main consideration when applying the method. At $2^{15}$ paths (used for Tables~\ref{tbl:vrfs-arithmetic}--\ref{tbl:vrfs-lookback}) we see a single kernel run take around $0.8$ms for the Brownian bridge construction method and $0.2$ms for the others. The basic CPU implementation at $2^{15}$ paths takes ${\sim}150$ms. As we move up to $2^{19}$ paths, QMC+BB-CPW requires $12$--$13$ms per kernel call and the other methods around $3$ms.

\subsection{Applicability and design}

Although only applied to three types of option, our method can be implemented for many types of options --- both vanilla and exotic. This allows for a single algorithm to be applied to a large set of the products an institution may work with and reduces the need for many distinct methods that depend on the option type, whilst also achieving a higher accuracy. For example, estimating the gamma of many option types is not possible through pathwise alone and so an existing solution would be to apply the likelihood ratio in conjunction with pathwise. Any variant of QMC-CPW is able to calculate gamma estimates so broadens the range of products that an institution can handle with much less overhead.

The templatised design of the simulation also allows other option types to be added easily, including those with multiple underlying assets. A redefinition of the path simulation and payoffs/Greeks formula for each type is all that is needed. We are also able to pull out Brownian bridge construction such that products can ingest the increments directly rather than the random normal variables $Z_i$.

One of the current limitations with the design is the lack of dynamic memory allocation, which would allow us to further encapsulate different product types and have finer-grained control over simulation. In Savine's book~\cite{savine2018aad}, a dynamic framework is presented in which options with a varying number of required random normal variables for simulation, can all follow the same path through the program. Our implementation has a fixed number of random numbers to generate at compile time and as such each product is required to take in all of those variables. This design was noted but the added difficulty of dealing with objects containing virtual functions in CUDA was seen as too far aside for the main objective of combining the QMC-CPW method and the parallel performance of the GPU.
\section{Conclusion} \label{sec:Conclusion}

We have presented a new efficient approach for calculating the Greeks of exotic options on the GPU. The Quasi-Monte Carlo Conditional Pathwise method developed by Zhang and Wang~\cite{ZhangConditionalQuasiMonteCarloMethod} allows for smoothing of the integrands which Quasi-Monte Carlo methods take advantage of to efficiently estimate the Greeks.

Our implementation uses the highly parallel nature of GPUs to efficiently implement the Quasi-Monte Carlo simulation such that our solution is hundreds of times faster than a serial CPU implementation. As a variance reduction technique, Brownian bridge construction is used in conjunction with the CPW estimates to further reduce the error in our Greek estimates. We show that our implementation, QMC+BB-CPW, produces estimates with VRFs in the hundreds of thousands and even up to $1.0 \times 10^{18}$ when compared to traditional methods such as the Likelihood Ratio method. When compared to other simulation methods such as MC+AV-CPW, our method outperforms for almost all Greek estimates of arithmetic Asian, binary Asian and lookback options over a range of strike prices.

Whilst the results obtained are more than satisfactory, we do not achieve VRFs of the same magnitude as in~\cite{ZhangConditionalQuasiMonteCarloMethod}. This is likely due to their implementation using Gradient Principle Component Analysis as a variance reduction technique which reduces the effective dimension, allowing Quasi-Monte Carlo methods to be even more efficient.

\subsubsection{Improved VRFs}

To achieve larger VRFs we could implement further variance reduction techniques such as (Gradient) Principle Component Analysis. Techniques such as this help Quasi-Monte Carlo methods to more efficiently estimate integrals as they typically reduce the effective dimension, which in the finance setting can be extremely useful due to the high dimensionaility of many problems.

\subsubsection{Towards an industry-grade implementation}

An industry-grade software solution could include improvements such as interfacing out the volatility model and other random number generators, and adding more option types and other financial products. 

Adding more flexibility to the software would also be a key requirement. In part this could be achieved by using dynamic data structures in kernels. Unified memory could enable us to freely, efficiently, and easily move objects from host to device and vice-versa, but could also restrict the software to GPU-only uses.

An industry-grade software product could wrap our implementation inside a microservice for calculating Greeks within a larger risk management system. It would receive parameters such as stock price, implied volatility and the expiration dates from the input bus, process these values to produce estimates for Greeks, and publish them to other microservices.

\paragraph{Acknowledgements} We would like to thank Prof.~Mike Giles (Mathematical Institute, University of Oxford) for many constructive comments and suggestions.

\bibliographystyle{alpha}

\appendix

\section{GPU and CUDA specifications} \label{app:gpu-cuda-specs}

CUDA toolkit version 11.2.1 was used for all of the software in this project. The online documentation for this version is available at \href{https://docs.nvidia.com/cuda/archive/11.2.1/}{https://docs.nvidia.com/cuda/archive/11.2.1/}.

All simulations were ran on a single Tesla T4 GPU which has the Turing architecture with compute capability $7.5$. The general information for the device is listed below:

\begin{table}[!h]
\centering
\begin{tabular}{c c}
    \textbf{General Information} & \\
    Name & Tesla T4 \\
    Compute Capability & $7.5$ \\
    Clock Rate (Hz) & $1590000$ \\
    Device Copy Overlap & Enabled \\
    Kernel Execution Timeout & Disabled \\ \\
    \textbf{Memory Information} & \\
    Global Memory & 15843721216 \\
    Constant Memory & 65536 \\
    Max Memory Pitch & 2147483647 \\
    Texture Alignment & 512 \\ \\
    \textbf{Multiprocessor Information} & \\
    Multiprocessor Count & 40 \\
    Shared Memory per MP & 49152 \\
    Registers per MP & 65536 \\
    Threads in Ward & 32 \\
    Max Threads per Block & 1024 \\
    Max Thread Dimensions & (1024, 1024, 64) \\
    Max Grid Dimensions & (2147483647, 65536, 65535) \\
\end{tabular}
\caption{Tesla T4 specifications. Memory values are given in bytes.}
\label{tbl:TeslaT4Specs}
\end{table}

\end{document}